\documentclass[twocolumn,showpacs,showkeys,preprintnumbers,superscriptaddress,amsmath,floatfix,amssymb,secnumarabic]{revtex4}
\usepackage[colorlinks=true]{hyperref}
\usepackage{graphicx}

\newcommand{\comment}[1]{}

\newcommand{\lr}[1]{ \left( #1 \right) }
\newcommand{\lrs}[1]{ \left[ #1 \right] }
\newcommand{\lrc}[1]{ \left\{ #1 \right\} }
\newcommand{\vev}[1]{ \langle \, #1 \, \rangle }

\newcommand{\tr}{ {\rm Tr} \, }

\newcommand{\expa}[1]{ \exp{\left( #1 \right)} }

\begin{document}
\sloppy
\preprint{ITEP-LAT/2010-07}

%\logo at the end of the title
\title{Schwinger-Dyson equations in large-$N$ quantum field theories and nonlinear random processes%\logo
}

\author{P. V. Buividovich}
\email{buividovich@itep.ru}
\affiliation{ITEP, 117218 Russia, Moscow, B. Cheremushkinskaya str. 25}
\affiliation{JINR, 141980 Russia, Moscow region, Dubna, Joliot-Curie str. 6}

\date{December 27, 2010}
\begin{abstract}
 We propose a stochastic method for solving Schwinger-Dyson equations in large-$N$ quantum field theories. Expectation values of single-trace operators are sampled by stationary probability distributions of the so-called nonlinear random processes. The set of all histories of such processes corresponds to the set of all planar diagrams in the perturbative expansions of the expectation values of singlet operators. We illustrate the method on the examples of the matrix-valued scalar field theory and the Weingarten model of random planar surfaces on the lattice. For theories with compact field variables, such as sigma-models or non-Abelian lattice gauge theories, the method does not converge in the physically most interesting weak-coupling limit. In this case one can absorb the divergences into a self-consistent redefinition of expansion parameters. Stochastic solution of the self-consistency conditions can be implemented as a ``memory'' of the random process, so that some parameters of the process are estimated from its previous history. We illustrate this idea on the example of two-dimensional $O\lr{N}$ sigma-model. Extension to non-Abelian lattice gauge theories is discussed.
\end{abstract}

\maketitle

\section*{Introduction}
\label{sec:introduction}

 Modern lattice QCD simulations are mostly based on direct evaluation of the path integral of the theory. Such approach, while being very general and efficient for many applications, suffers from a number of problems, most notable of which are the sign problem at finite chemical potential, critical slowing down at small quark masses and large finite-volume effects as well as small signal-to-noise ratio in the analysis of excited states. These problems are inherent to standard Monte-Carlo simulations and cannot be efficiently solved by simply increasing the computation power, since the required computing time quickly increases (in the worst cases, exponentially) with the required precision. Such situation makes it tempting to devise alternative simulation algorithms for non-Abelian lattice gauge theories.

 One of the efficient alternative numerical methods is the so-called Diagrammatic Monte-Carlo, a method based on stochastic summation of all the terms in the strong- or weak-coupling expansion of the observable of interest \cite{Prokofev:98:1, Wolff:09:1}. Such a method in some cases allows one to reduce or avoid completely the sign problem in the original path integral, and does not suffer from finite-volume effects. Furthermore, one can construct algorithms which yield particular correlation functions in terms of probability distributions of some random variables, which greatly facilitates the analysis of excited states \cite{Prokofev:98:1, Wolff:09:1}. This is the idea of the ``worm'' algorithm by Prokof'ev and Svistunov \cite{Prokofev:98:1}, in which the probability distribution of the positions $x$, $y$ of ``head'' and the ``tail'' of the worm yields the two-point Green function $G\lr{x, y}$. Diagrammatic Monte-Carlo and the ``worm'' algorithm have been successfully applied to a number of statistical models with discrete symmetry groups such as the Ising model, the XY model and unitary Fermi gas and showed practically no critical slowing down near quantum phase transitions.

 However, application of such methods to lattice field theories with continuous field variables (such as two-dimensional $O\lr{N}$ and $CP\lr{N}$ sigma-models, Abelian gauge theories and the $\phi^4$ theory) resulted so far in quite complicated and model-dependent algorithms \cite{Wolff:09:1}. A generalization of such algorithms to $SU\lr{N}$ sigma-models or to non-Abelian gauge theories is still not found. These algorithms are in essence based on the strong-coupling expansion, and while their applicability is not limited by the strong-coupling regime, one can expect that algorithms based on the weak-coupling expansion might show better performance near the continuum limit.

 Typically, the weak-coupling expansion in such lattice theories is either quite complicated or non-convergent. Up to now, divergent behavior of the weak-coupling perturbative expansions strongly limits the applicability of Diagrammatic Monte-Carlo to field theories with continuous field variables. In a recent paper \cite{Prokofev:10:1} a method was proposed to construct convergent series which approximate the non-analytic path integrals with desired precision. This method, however, is difficult to generalize to physically interesting field theories such as non-Abelian lattice gauge theories.

 Another way to obtain convergent series while preserving important physical properties of the theory is to sum over diagrams with certain topology only. This corresponds to the large-$N$ limit in quantum field theories and matrix models, that is, the limit of infinite dimensionality of an internal symmetry group, such as $O\lr{N}$ or $SU\lr{N}$. For such theories, each Feynman diagram acquires a factor $N^{\chi}$, where $\chi$ is the Euler character of this diagram \cite{tHooft:75:1}. In the limit $N \rightarrow \infty$, the contribution of planar diagrams with $\chi = 2$ dominates, and the sum over all planar diagrams typically has a finite convergence radius \cite{Koplik:77:1, Brezin:78:1}.

 In this paper we describe a stochastic method for summing over all planar diagrams in large-$N$ quantum field theories. The method is based on stochastic solution of Schwinger-Dyson equations, so that the correlators of field variables are obtained as stationary probability distributions of certain random variables. In this way we implement the idea of importance sampling, so that numerically small observables correspond to unlikely events. These probability distributions are sampled by the so-called nonlinear random processes. In contrast to conventional Markov chains, stationary probability distributions of such random processes satisfy nonlinear equations, and hence they can be called ``nonlinear random processes'' or ``nonlinear Markov chains'' in the terminology of \cite{Etessami:05:1, Frank:04:1}. Factorization of single-trace operators in the large-$N$ limit of quantum field theories corresponds to the phenomena of ``chaos propagation'' in random processes \cite{KacProbability}.

 While in the diagrammatic Monte-Carlo and in the ``worm'' algorithm the diagrams are stored in computer memory as a whole and are updated in such a way that the detailed balance condition is satisfied at each step, the method described in this paper works only with external lines. In contrast to the standard Metropolis algorithm, one should not know explicitly the weight of each diagram, and the transition probabilities do not satisfy any detailed balance condition. Unlike the quite popular ``numerical functional methods'' in continuum gauge theories (see \cite{Pawlowski:07:1} for a review), the proposed method does not require any truncation of the hierarchy of Schwinger-Dyson equations, and work only with singlet operators w.r.t. the internal symmetry group. Another distinct feature is that the computational complexity of the method does not depend on $N$, while the standard Monte-Carlo, the functional methods and the ``worm'' algorithm all require infinite computational resources in the limit $N \rightarrow \infty$. This feature might be advantageous for numerical checks of the predictions of the holographic models which are dual to large-$N$ quantum field theories \cite{Polyakov:99:1}.

 In Section \ref{sec:SDeq_general} we analyze the general structure of Schwinger-Dyson equations in large-$N$ quantum field theories on the example of a scalar matrix-valued field theory. When large-$N$ factorization is taken into account, Schwinger-Dyson equations become nonlinear equations with infinitely many unknowns. In Section \ref{sec:recursive_process} we describe nonlinear random processes of recursive type \cite{Etessami:05:1} which can be used to stochastically solve such equations. In Section \ref{sec:SDs_stochastic_solution} we apply such random processes to solve Schwinger-Dyson equations in several large-$N$ theories. In Subsection \ref{subsec:phi4_general} we consider the scalar matrix-valued field theory, for which the perturbative expansion yields the conventional Feynman diagrams in momentum space. In Subsection \ref{subsec:matrix_model} this solution is compared with the exact solution of the simplest quantum field theory in zero dimensions, that is, the Hermitian matrix model \cite{Brezin:78:1}. The convergence of such solution and the strength of the sign problem is discussed. In Subsection \ref{subsec:weingarten} we consider the Weingarten model \cite{Weingarten:80:1, Eguchi:82:3} and demonstrate how the proposed method can be used to simulate random surfaces on the hypercubic lattice. In this case, our method reproduces an ensemble of open, rather than closed, random surfaces, with critical behavior which is quite different from those of the closed planar random surfaces. Since the structure of Schwinger-Dyson equations in the Weingarten model is similar to the loop equations in large-$N$ non-Abelian lattice gauge theories \cite{Migdal:81:1}, studying this model might be helpful for further extensions of the present approach to non-Abelian gauge theories.

 While the method described in Section \ref{sec:recursive_process} works well for non-compact field variables, for field theories with compact field variables, such as nonlinear sigma-models or non-Abelian lattice gauge theories, a straightforward stochastic interpretation of Schwinger-Dyson equations is only possible in the strong coupling limit. In the weak-coupling limit one expects the field correlators to contain both the perturbative part in the coupling constant $g$ as well as nonperturbative corrections of the form $\expa{-c/g^2}$ with some constant $c$. Moreover, perturbative expansion in powers of $g$ typically results in asymptotic series, and nonperturbative corrections appear as a result of resummation of such series \cite{Parisi:77:1}.

 In Section \ref{sec:rps_with_mem} we show how such nonperturbative corrections can be taken into account by a further relaxation of the Markov property of the random process. The basic idea is to absorb the divergent part of the series into a self-consistent redefinition of the expansion parameter. These redefined parameters play the role of nonperturbative ``condensates'' \cite{Parisi:77:1, Kazakov:94:1}. It turns out that the redefined expansion parameters can be estimated with increasing precision from the previous history of the random process which solves the Schwinger-Dyson equations, thus leading to the emergence of the ``memory'' of the random process. The approach of the redefined parameters to their self-consistent values is reminiscent somehow of the renormalization-group flow \cite{Pawlowski:07:1}. Such dependence on the previous history makes the random process essentially non-Markovian, so that the stationary probability distribution also satisfies some nonlinear equation.

 We illustrate this idea on the example of $O\lr{N}$ sigma-model in two dimensions, which is equivalent to a bosonic random walk with a self-consistent mass. Random process which simulates this model has the ``memory'' but no ``recursive'' structure. Presumably, in order to sum up both perturbative and non-perturbative corrections which arise in non-Abelian lattice gauge theories or $U\lr{N}$ sigma-models, one should devise the ``recursive'' nonlinear random process (which would sum up perturbative corrections) with memory (which would generate nonperturbative quantities in a renormalization-group-like way).

 Finally, in the concluding Section we summarize the present work and discuss its extension to non-Abelian lattice gauge theories in the limit of large $N$.

\section{General structure of Schwinger-Dyson equations for large-$N$ quantum field theories}
\label{sec:SDeq_general}

 In order to analyze the general structure of Schwinger-Dyson equations for large-$N$ quantum field theories, let us first consider the theory of a hermitian $N\times N$ matrix-valued field $\phi\lr{x}$ with the following Lagrangian:
\begin{eqnarray}
\label{phi4_field_action}
\mathcal{L}\lrs{\phi\lr{x}} = N \tr \phi\lr{x} \, \lr{m^2 - \Delta} \, \phi\lr{x} + \frac{N \lambda}{4} \tr \phi^{4}\lr{x}  .
\end{eqnarray}
This theory is most convenient to illustrate the method described in this paper, since its perturbative expansion leads to conventional Feynman diagrams in the momentum space. Since this theory should be somehow regularized, let us assume from the very beginning that the action (\ref{phi4_field_action}) is defined on the Euclidean hypercubic $D$-dimensional lattice with total volume $V$ in lattice units. Thus, the coordinates $x$ take integer values and $\Delta$ is the lattice Laplacian (for definiteness, with periodic boundary conditions).

\begin{figure}
  \includegraphics[width=6cm]{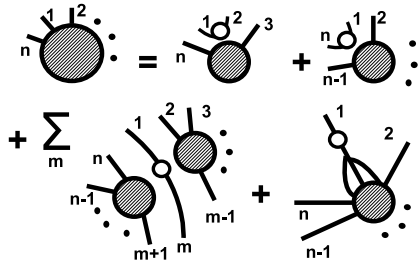}\\
  \caption{Schematic illustration of the structure of the Schwinger-Dyson equations (\ref{phi4_SDs_momentum}). Dashed blobs denote the Green functions $G\lr{k_1, \ldots, k_n}$ and empty blobs denote the free propagator $G_0\lr{k}$.}
  \label{fig:phi4_SDs_general}
\end{figure}

 Schwinger-Dyson equations for a theory with the action (\ref{phi4_field_action}) read \cite{Arefeva:81:1}:
\begin{widetext}
\begin{eqnarray}
\label{phi4_SDs_original_n2}
\lr{m^2 - \Delta_1} \, G\lr{x_1, x_2} = \delta\lr{x_1, x_2} + \lambda \, G\lr{x_1, x_1, x_1, x_2}  , \\
\label{phi4_SDs_original}
\lr{m^2 - \Delta_1} \, G\lr{x_1, \ldots, x_n} =
\delta\lr{x_1, x_2} \, G\lr{x_3, \ldots, x_n}
 + \nonumber \\ +
\delta\lr{x_1, x_n} \, G\lr{x_2, \ldots, x_{n-1}}
 +
\sum \limits_{m=3}^{n-1} \delta\lr{x_1, x_m} \, G\lr{x_2, \ldots, x_{m-1}} \, G\lr{x_{m+1}, \ldots, x_n}
 + \nonumber \\ +
\lambda \, G\lr{x_1, x_1, x_1, x_2, \ldots, x_n}  ,
\quad \quad n > 2,
\end{eqnarray}
where the single-trace correlators are $G\lr{x_1, \ldots, x_n} = \vev{ \frac{1}{N} \, \tr\lr{ \phi\lr{x_1} \ldots \phi\lr{x_n} } }$, $\Delta_1$ is the Laplacian acting on $x_1$ and we have already taken into account the factorization property in the limit $N \rightarrow \infty$ \cite{tHooft:75:1}:
\begin{eqnarray}
\label{largeNfactorization}
\vev{\frac{1}{N} \tr\lr{\phi\lr{x_1} \ldots \phi\lr{x_n} } \, \frac{1}{N} \tr\lr{\phi\lr{y_1} \ldots \phi\lr{y_m} } }
= \nonumber \\ =
\vev{\frac{1}{N} \tr\lr{\phi\lr{x_1} \ldots \phi\lr{x_n} }} \, \vev{\frac{1}{N} \tr\lr{\phi\lr{y_1} \ldots \phi\lr{y_m} } } + O\lr{\frac{1}{N^2}}  .
\end{eqnarray}
These equations hold for any argument of the correlators, but the resulting system is redundant, and it is sufficient to consider only those Schwinger-Dyson equations which were obtained by the variation of the fields at $x_1$.

 It is convenient now to go to the momentum representation, introducing the Green functions in momentum space $G\lr{k_1, \ldots, k_n} = \sum \limits_{x_1} \ldots \sum \limits_{x_n} \, \expa{i \sum \limits_{m} k_m \cdot x_m} \, G\lr{x_1, \ldots, x_n}$. In order to keep all expressions as symmetric as possible, we do not separate the factor $\delta\lr{\sum \limits_{m} k_m}$ in $G\lr{k_1, \ldots, k_n}$ explicitly. This condition will be automatically satisfied by the nonlinear random process which we describe in Subsection \ref{subsec:phi4_general}. The equations (\ref{phi4_SDs_original_n2}), (\ref{phi4_SDs_original}) in the momentum representation are:
\begin{eqnarray}
\label{phi4_SDs_momentum_n2}
G\lr{k_1, k_2} = G_0\lr{k_1} \, V \, \delta\lr{k_1 + k_2}
 + \nonumber \\ +
G_0\lr{k_1} \, \frac{\lambda}{V^2} \, \sum \limits_{q_1, q_2, q_3} \delta\lr{k_1 - q_1 - q_2 - q_3} \, G\lr{q_1, q_2, q_3, k_2}  , \\
\label{phi4_SDs_momentum}
G\lr{k_1, \ldots, k_n}
=
G_0\lr{k_1} \, \sum \limits_{m=3}^{n-1} \delta\lr{k_1 + k_m} \, V \, G\lr{k_2, \ldots, k_{m-1}} \, G\lr{k_{m+1}, \ldots, k_n}
+ \nonumber\\ +
G_0\lr{k_1} \, \delta\lr{k_1 + k_2}\, V\, G\lr{k_3, \ldots, k_n}
+
G_0\lr{k_1} \, \delta\lr{k_1 + k_n}\, V\, G\lr{k_2, \ldots, k_{n-1}}
+ \nonumber\\ +
G_0\lr{k_1} \, \frac{\lambda}{V^2} \, \sum \limits_{q_1, q_2, q_3} \delta\lr{k_1 - q_1 - q_2 - q_3} \, G\lr{q_1, q_2, q_3, k_2, \ldots, k_n} ,
\end{eqnarray}
where $G_0\lr{k} = \lr{m^2 + \sum \limits_{\mu} 4 \sin^2\lr{k_{\mu}/2}}^{-1}$ is the free scalar propagator on the hypercubic lattice. All momenta are assumed to lie in the first Brillouin zone $-\pi \le k_{\mu} \le \pi$ and are added modulo $2 \pi$. The structure of these equations is schematically illustrated on Fig. \ref{fig:phi4_SDs_general}, where dashed blobs denote the Green functions $G\lr{k_1, \ldots, k_n}$ and empty blobs denote $G_0\lr{k}$.
\end{widetext}

 Thus we have obtained an infinite system of quadratic functional equations for the set of functions $G\lr{k_1, \ldots, k_n}$ with $n = 2, 4, \ldots$. Such structure is common for large-$N$ quantum field theories: Schwinger-Dyson equations are quadratic equations for infinite set of unknown variables. In the case of scalar matrix field theory considered here, the unknown variables are the functions of the sequences of momenta $\lrc{k_1, \ldots, k_n}$ for any even $n \ge 2$. In the case of lattice gauge theories or string theories Schwinger-Dyson equations are most naturally formulated in terms of the Wilson loops, which are the functions defined on the discrete space of closed loops on the lattice \cite{Migdal:81:1, Weingarten:80:1, Eguchi:82:3}. In this case, the equations are also quadratic w.r.t. the Wilson loops. For $O\lr{N}$ sigma-model, Schwinger-Dyson equations are also quadratic equations which involve only the two-point function (see Section \ref{sec:rps_with_mem}).

 Typically, systems of equations with infinitely many unknowns can be efficiently solved by stochastic methods. It is advantageous to estimate the value of each unknown variable as a probability of observing some state of a random process. In this case the unknowns with numerically small values correspond to unlikely events, and the set of infinitely many unknown variables is automatically truncated to a set of unknowns with sufficiently large values. Such methods are well-known mainly in the context of kinetic equations \cite{KacProbability}. Recently they were also discussed in the context of probabilistic programming \cite{Etessami:05:1}. In the next Section we describe a discrete-time, discrete-space method of such type which is in our opinion most suitable for solving the Schwinger-Dyson equations in the large-$N$ limit.

\section{Stochastic solution of nonlinear equations by random processes of recursive type}
\label{sec:recursive_process}

 We consider nonlinear equations of the following form:
\begin{eqnarray}
\label{random_process_eq}
w\lr{x} = p_{c}\lr{x} + \sum \limits_{y} p_{e}\lr{x | y} w\lr{y}
 +  \nonumber \\ + %preprint
\sum \limits_{y_1, y_2} p_{j}\lr{x | y_1, y_2} w\lr{y_1} w\lr{y_2}  ,
\end{eqnarray}
where $x$, $y$, $y_1$, $y_2$ are the elements of some space $X$ and $\sum \limits_{x}$ with $x \in X$ denotes summation or integration over all the elements of this space. We also assume that the functions $p_{c}\lr{x}$, $p_{e}\lr{x|y}$ and $p_{j}\lr{x | y_1, y_2}$ satisfy the inequalities
\begin{eqnarray}
\label{probability_ineq}
 \sum \limits_{x} |p_{c}\lr{x}| + |p_{e}\lr{x|y_1}| + |p_{j}\lr{x | y_1, y_2}| < 1
\end{eqnarray}
for any $y_1$, $y_2$.

 We would like to find a stochastic process for which $w\lr{x}$ is proportional to the probability of the occurrence of the element $x$ in some configuration space. Obviously, ordinary Markov process with configuration space $X$ cannot solve such a problem, since stationary distributions of Markov processes obey linear equations. In order to solve the nonlinear equation (\ref{random_process_eq}), one can, for example, extend somehow the configuration space. Extensions of Markov processes with stationary probability distributions which obey nonlinear equations have been considered recently in \cite{Etessami:05:1, Frank:04:1}. In this Section we concentrate on random processes similar to recursive Markov chains of \cite{Etessami:05:1}. The basic idea is that at any time one can leave the current chain and start a new one, then returning back to the old chain at some time. The initial state of a newly created chain depends on the states of older chains. Thus one has not a single Markov chain, but rather an infinite stack of chains. The random process which we describe below will be similar to these recursive Markov chains, but instead of referring to ``recursion'' we will explicitly introduce the underlying stack structure. Here we first consider the equations (\ref{random_process_eq}) with the coefficients $p_{c}\lr{x}$, $p_{e}\lr{x|y}$ and $p_{j}\lr{x | y_1, y_2}$ being all positive, and in Appendix \ref{sec:arbitrary_coefficients} we generalize to coefficients with arbitrary signs or complex phases.

 Consider an extended configuration space which consists of ordered sequences $\lrc{x_1, \ldots, x_n}$ for arbitrary $n \ge 1$, with $x_1, \ldots, x_n \in X$. It is illustrative to interpret such configuration space as a stack of elements of the space $X$, so that $x_n$ is at the top of the stack. The desired random process can be specified by the following prescriptions. At each discrete time step do one of the following:
\begin{description}
  \item[Create:] With probability $p_{c}\lr{x}$ create new element $x \in X$ and push it to the stack.
  \item[Evolve:] With probability $p_{e}\lr{x | y}$ pop the element $y$ from the stack and push the element $x$ to the stack.
  \item[Join:] With probability $p_{j}\lr{x | y_1, y_2}$ consecutively pop two elements $y_1$, $y_2$ from the stack and push a single element $x$ to the stack.
  \item[Restart:] With probability $1-\sum\limits_{x}\lr{ p_{c}\lr{x} + p_{e}\lr{x|y_1} + p_{j}\lr{x | y_1, y_2}}$, where $y_1$, $y_2$ are the two topmost elements in the stack, empty the stack and push a single element $x \in X$ into it, with probability distribution proportional to $p_{c}\lr{x}$.
\end{description}
The last action is also the procedure used to initialize the random process. The ``Evolve'' action is just the evolution of a single Markov chain at the top of the stack, with transition probabilities proportional to $p_{e}\lr{x | y}$. The condition (\ref{probability_ineq}) and the positivity requirement ensures that $p_{c}\lr{x}$, $p_{e}\lr{x | y}$ and $p_{j}\lr{x | y_1, y_2}$ can be interpreted as probabilities.

 Consider now an equation for the stationary probability distribution of such a Markov chain. It has a general form $p\lr{A} = \sum \limits_{B} P\lr{B \rightarrow A} \, p\lr{B}$, where $p\lr{A}$ is a stationary probability of the occurrence of a state $A$ and $P\lr{B \rightarrow A}$ is the transition probability between the states $B$ and $A$. Let $W\lr{x_1, \ldots, x_n}$ be the stationary probability to find the elements $x_1, \ldots, x_n$ in the stack. This probability distribution function is obviously normalized to unity:
$$\sum \limits_{n=1}^{\infty} \sum \limits_{x_1} \ldots \sum \limits_{x_n} W\lr{x_1, \ldots x_n} = 1 . $$
The equation for the stationary probability distribution in our case reads:
\begin{eqnarray}
\label{random_process_large_stationary_eq_n2}
W\lr{x_1}
 =
\mathcal{N}_c^{-1} \, p_c\lr{x_1} \, \xi_R
 +
\sum \limits_{y} p_e\lr{x_1 | y} \, W\lr{y}
 + \nonumber \\ + %preprint
\sum \limits_{y_1, y_2} p_j\lr{x_1 | y_1, y_2} \, W\lr{y_1, y_2}
\\ \label{random_process_large_stationary_eq}
W\lr{x_1, \ldots, x_n}
 =
p_c\lr{x_n} \, W\lr{x_1, \ldots, x_{n-1}}
 +  \nonumber \\ + %preprint
\sum \limits_{y} p_e\lr{x_n | y} \, W\lr{x_1, \ldots, x_{n-1}, y}
 + \nonumber \\ +
\sum \limits_{y_1, y_2} p_j\lr{x_n | y_1, y_2} \, W\lr{x_1, \ldots, x_{n-1}, y_1, y_2}
  ,  n > 1  ,
\end{eqnarray}
where
\begin{eqnarray}
\label{xi_R_def}
\xi_R = \sum \limits_{n} \sum \limits_{x_1} \ldots \sum \limits_{x_n}
\left(1 -  \right.
p_c\lr{x_n} \, W\lr{x_1, \ldots, x_{n-1}}
 - \nonumber \\ - %preprint
\sum \limits_{y} p_e\lr{x_n | y} \, W\lr{x_1, \ldots, x_{n-1}, y}
 - \nonumber \\ -
\left.
\sum \limits_{y_1, y_2} p_j\lr{x_n | y_1, y_2} \, W\lr{x_1, \ldots, x_{n-1}, y_1, y_2}
\right)
\end{eqnarray}
and $\mathcal{N}_c = \sum \limits_x p_c\lr{x}$. By a direct substitution one can check that there is a factorized solution for $W\lr{x_1, \ldots, x_n}$:
\begin{eqnarray}
\label{random_process_factorized_pd}
 W\lr{x_1, \ldots, x_n} = w_{0}\lr{x_1} \, w\lr{x_2} \, \ldots w\lr{x_n} ,
\end{eqnarray}
where $w\lr{x}$ obeys exactly the equation (\ref{random_process_eq}) and $w_0\lr{x}$ obeys the following inhomogeneous linear equation:
\begin{eqnarray}
\label{random_process_eq_w0}
 w_0\lr{x}
 =
\mathcal{N}_c^{-1} \, p_c\lr{x} \, \xi_R
 +
\sum \limits_{y} p_e\lr{x | y} \, w_0\lr{y}
 +  \nonumber \\ + %preprint
\sum \limits_{y_1, y_2} p_j\lr{x | y_1, y_2} \, w_0\lr{y_1} w\lr{y_2} .
\end{eqnarray}

 Thus, for any equation of the form (\ref{random_process_eq}) with positive coefficients which satisfy (\ref{probability_ineq}), there is a random process whose stationary distribution encodes the solution of this equation as in (\ref{random_process_factorized_pd}). The factorization of the stationary probability distribution of random processes with such an infinite configuration space is known as the ``propagation of chaos'' in random processes and was discovered for classical kinetic equations by McKean, Vlasov and Kac \cite{KacProbability}. Comparing the equation (\ref{random_process_eq}) with Schwinger-Dyson equations (\ref{phi4_SDs_original_n2}), (\ref{phi4_SDs_original}), (\ref{phi4_SDs_momentum_n2}) and (\ref{phi4_SDs_momentum}), we conclude that this property corresponds to the factorization of single-trace operators in large-$N$ quantum field theories. It is interesting that time reversal of the random process described above leads to the so-called branching random process \cite{KacProbability}, which has quite different properties. This is due to the fact that for such random processes there is no detailed balance condition, and hence no time reversal symmetry. We do not consider here a subtle mathematical question of the existence of solutions to equation (\ref{random_process_eq}), since in our case it is ensured by the physical applications of this equation.

 Finally, let us describe a practical procedure for finding $w\lr{x}$ by simulating the random process described above. By standard statistical methods, one should sample the probability distribution $p\lr{x_n}$ of the topmost element in the stack (provided there is more than one element in it, otherwise we estimate $w_0\lr{x}$ rather than $w\lr{x}$, see (\ref{random_process_factorized_pd})). From (\ref{random_process_factorized_pd}), we get $p\lr{x_n} = s^{-1} w\lr{x_n}$, with $s = \sum \limits_{x} w\lr{x}$. It should be stressed that $w\lr{x}$ is not normalized to unity, but rather satisfies the inequality $\sum \limits_{x} w\lr{x} = s < 1$. The value of the normalization constant $s$ can be also easily found numerically, since the probability to find $n$ elements in the stack decreases as $s^n$ for $n > 1$.

\section{Stochastic solution of Schwinger-Dyson equations by recursive random processes}
\label{sec:SDs_stochastic_solution}

\subsection{Scalar matrix field theory}
\label{subsec:phi4_general}

 After presenting the general method in Section \ref{sec:recursive_process}, we are ready to describe a stochastic numerical solution of Schwinger-Dyson equations (\ref{phi4_SDs_momentum_n2}), (\ref{phi4_SDs_momentum}). For simplicity, let us assume that the coupling constant $\lambda$ in (\ref{phi4_field_action}) is negative. This allows us to apply directly the results of Section \ref{sec:recursive_process}, where all the coefficients in (\ref{random_process_eq}) are assumed to be positive. In the case of positive $\lambda$, additional sign variables for each sequence of momenta can be easily introduced following Appendix \ref{sec:arbitrary_coefficients}. This will be done in the next Subsection for the Hermitian matrix model. Note that while at finite $N$ the theory with negative coupling constant is not defined and the correlators are non-analytic in $\lambda$ \cite{Prokofev:10:1}, in the leading order in $N$ perturbative series converge even when the coupling is negative, but not exceeding some critical value \cite{Brezin:78:1}. Correspondingly, in the planar approximation the correlators are analytic in $\lambda$.

 The space $X$ in (\ref{random_process_eq}) should be the space of ordered sequences (of any size) of momenta $\lrc{k_1, \ldots, k_n}$ , correspondingly, the extended configuration space is a stack which contains such sequences. It is convenient also to introduce two normalization constants $\mathcal{N}$ and $c$, so that the functions $w\lr{k_1, \ldots k_n}$ which will be estimated stochastically are defined as
\begin{eqnarray}
\label{phi4_SDs_momentum_PDF}
G\lr{k_1, \ldots, k_n} = \mathcal{N} \, V^{n} \, c^{n-2} \, w\lr{k_1, \ldots, k_n}  ,
\end{eqnarray}
where $V$ is again the total volume of space. The constant $c$ can be thought of as the renormalization constant for the one-particle wave functions, and $\mathcal{N}$ - as the overall wavefunction normalization.

 In terms of the functions $w\lr{k_1, \ldots, k_n}$ the Schwinger-Dyson equations (\ref{phi4_SDs_momentum_n2}) and (\ref{phi4_SDs_momentum}) read:
\begin{widetext}
\begin{eqnarray}
\label{phi4_SDs_momentum_n2_rescaled}
w\lr{k_1, k_2}
=
G_0\lr{k_1} \, \mathcal{N}^{-1} \frac{\delta\lr{k_1 + k_2}}{V}
+  \nonumber \\ +
G_0\lr{k_1} \, \lambda c^2 \, \sum \limits_{q_1, q_2, q_3} \delta\lr{k_1 - q_1 - q_2 - q_3} \, w\lr{q_1, q_2, q_3, k_2}  ,
\\
\label{phi4_SDs_momentum_rescaled}
w\lr{k_1, \ldots, k_n}
\, = \,
G_0\lr{k_1} \, c^{-2} \frac{\delta\lr{k_1 + k_2}}{V} \, w\lr{k_3, \ldots, k_n}
+ \nonumber \\ +
G_0\lr{k_1} \, c^{-2} \frac{\delta\lr{k_1 + k_n}}{V} \, w\lr{k_2, \ldots, k_{n-1}}
+ \nonumber \\ +
G_0\lr{k_1} \mathcal{N} \, c^{-4} \sum \limits_{m=3}^{n-1} \frac{\delta\lr{k_1 + k_m}}{V} \, w\lr{k_2, \ldots, k_{m-1}} \, w\lr{k_{m+1}, \ldots, k_n}
- \nonumber\\ -
G_0\lr{k_1} \, \lambda c^2 \, \sum \limits_{q_1, q_2, q_3} \delta\lr{k_1 - q_1 - q_2 - q_3} \, w\lr{q_1, q_2, q_3, k_2, \ldots, k_n}  .
\end{eqnarray}
\end{widetext}

 Comparing the Schwinger-Dyson equations (\ref{phi4_SDs_momentum_n2_rescaled}), (\ref{phi4_SDs_momentum_rescaled}) with the general equation (\ref{random_process_eq}), we arrive at the random process which stochastically solves these equations. This random process is specified by the following probabilistic choice of actions at each discrete time step:
\begin{description}
  \item[Create:] With probability $G_0\lr{k} \lr{\mathcal{N} V}^{-1}$ push a new sequence of momenta $\lrc{k, -k}$ to the stack.
  \item[Add:] With probability $G_0\lr{k} c^{-2}/V$ modify the topmost sequence of momenta $\lrc{k_1, \ldots, k_n}$ in the stack by adding a pair of momenta $\lrc{k, -k}$ either as $\lrc{k, k_1, \ldots, k_n, -k}$ or $\lrc{k, -k, k_1, \ldots, k_n}$.
  \item[Create vertex:] With probability $ |\lambda| \, G_0\lr{q_1 + q_2 + q_3} c^2$ replace the topmost sequence $\lrc{q_1, q_2, q_3, k_2, \ldots, k_n}$ in the stack by $\lrc{q_1 + q_2 + q_3, k_2, \ldots, k_n}$. This action can only be performed if the topmost sequence contains more than two elements.
  \item[Join:] With probability $G_0\lr{k} \mathcal{N}\, c^{-4} / V$ pop the two sequences $\lrc{k_1, \ldots, k_n}$, $\lrc{q_1, \ldots, q_m}$ from the stack (provided there are more than two elements in it) and join them into a single sequence as $\lrc{k_1, \ldots, k_n, k, q_1, \ldots, q_n, -k}$. Push the result to the stack.
  \item[Restart:] Otherwise restart with a stack containing a sequence $\lrc{k,-k}$, $k$ being distributed with the probability proportional to $G_0\lr{k}$
\end{description}
Since the momenta are always added to the stack in pairs which sum up to zero, for all sequences in the stack the total sum of all momenta in the sequence is always zero. The $V^{-1}$ factors in (\ref{phi4_SDs_momentum_n2_rescaled}), (\ref{phi4_SDs_momentum_rescaled}) ensure that the probability distributions of the newly created momenta can be normalized to unity.

 Let us check whether the inequalities (\ref{probability_ineq}) are satisfied for such a process, that is, whether the total probability of all possible actions does not exceed unity. For the free propagator $G_0\lr{k}$ one has the inequalities $G_{0}\lr{k} < 1/m^2$ and $\sum \limits_{k} G_{0}\lr{k} < V/m^2$. The total probability of all possible actions can be then estimated as $\lr{\mathcal{N}^{-1} + c^{-2} + |\lambda| c^2 + \mathcal{N}c^{-4}}/m^2$. Clearly, for sufficiently small $|\lambda|$ this estimate can be always made smaller than unity by increasing $c$ and $\mathcal{N}$. In Subsections \ref{subsec:matrix_model} and \ref{subsec:weingarten} we will analyze such bounds on coupling constants in more details for Hermitian matrix model and for the Weingarten model.

 Since the constructed process involves no permutations, one can trace the history of each momenta in the stack - from creation to joining into a vertex or a ``Restart operation''. By drawing all the momenta in stack as points on the vertical lines of some two-dimensional grid and connecting the corresponding points along the horizontal lines, all planar diagrams of the theory (with an arbitrary number of external lines)  can be obtained. Note also that the number of vertices in planar diagrams drawn by this random process cannot exceed the number of time steps from the previous ``Restart'' action. Thus in order to maximize the mean order of diagrams which are summed up in some fixed number of time steps, it is advantageous to maximally reduce the rate of ``Restart'' events, that is, to saturate the inequalities (\ref{probability_ineq}).

 One could also try to devise a random process which would solve the Schwinger-Dyson equations (\ref{phi4_SDs_original_n2}), (\ref{phi4_SDs_original}) directly in physical space-time, rather than in the momentum space. The configuration space of such a process would be the stack of sequences of points $\lrc{x_1, \ldots, x_n}$. As compared to the algorithm in the momentum space, there would be an additional choice of moving the last point $x_n$ in the topmost sequence to adjacent lattice sites, with the probability proportional to the hopping parameter $\kappa = \lr{2 D + m^2}^{-1}$. This would correspond to drawing the worldlines of virtual and real particles by bosonic random walks. Interestingly, such worldlines can be mapped onto the string worldsheets in simplicial string theory \cite{Akhmedov:04:1}.  As well, the creation of a new interaction vertex would only be possible if three such random walks would intersect in one point. However, this is an unlikely event, with probability going to zero in the continuum limit. Thus, solving the Schwinger-Dyson equations directly in the coordinate representation would lead to a less efficient numerical algorithm.

 Note that for the theory (\ref{phi4_field_action}) at finite $N$ the Schwinger-Dyson equations are linear equations, which are, however, defined on much larger functional space: the set of unknown functions includes also expectation values of multi-trace operators, such as $\vev{\tr\lr{\phi\lr{x_1} \ldots \phi\lr{x_n}} \, \tr\lr{\phi\lr{y_1} \ldots \phi\lr{y_m}}}$. One can try to solve these linear equations by interpreting them as the equations for the stationary probability distribution of a Markov process. The configuration space of such a process should be a space of sequences of the form $\lrc{\lrc{x_1, \ldots, x_n}, \ldots, \lrc{y_1, \ldots, y_m}}$, thus encoding the expectation values of all multi-trace operators. However, such a straightforward procedure leads to non-normalizable transition probabilities, indicating that the series which one tries to sum up are divergent. Only when the terms subleading in $1/N$ are omitted from the Schwinger-Dyson equations, they can be interpreted as stochastic equations. At the same time, we obtain the Markov process on the extended configuration space described in Section \ref{sec:recursive_process}, which we interpret as the stack of sequences. The property of the ``propagation of chaos'' \cite{KacProbability} ensures large-$N$ factorization of single-trace operators (see equation (\ref{random_process_factorized_pd})). We are thus led to the random process of recursive type \cite{Etessami:05:1}.

\subsection{Hermitian matrix model}
\label{subsec:matrix_model}

 To check the considerations of the previous Subsection, let us consider the theory (\ref{phi4_field_action}) in zero dimensions, that is, the hermitian matrix model with the following partition function:
\begin{eqnarray}
\label{matrix_model_def}
\mathcal{Z}\lr{\lambda} = \int \prod \limits_{i,j} d \phi_{ij} \,
\expa{ -N/2 \, \tr\phi^2 + \frac{\lambda N}{4}\, \tr\phi^4 }  .
\end{eqnarray}
The Green functions now depend only on one integer $n$: $G_n \equiv G\lr{n} = \vev{\frac{1}{N} \, \tr \phi^{2 n}}$. The Schwinger-Dyson equations (\ref{phi4_SDs_original_n2}), (\ref{phi4_SDs_original}) also take a very simple form:
\begin{eqnarray}
\label{matrix_model_SDs}
 G_1 = 1 + \lambda G_2,
\nonumber \\
 G_n = 2 G_{n-1} + \sum \limits_{m = 1}^{n-2} G_m \, G_{n-m-1} + \lambda G_{n+1}, \quad n > 1.
\end{eqnarray}
Here we will assume that the coupling constant $\lambda$ can be both positive and negative, in order to illustrate the method described in Appendix \ref{sec:arbitrary_coefficients}. Let us again define the ``renormalized'' Green functions $w_n$ as $G_n = \mathcal{N} c^{n-1} \, w_n$. In the case of arbitrary sign of $\lambda$, the configuration space of the random process should be the stack which contains integer positive numbers and additional sign variables. Following Appendix \ref{sec:arbitrary_coefficients}, we introduce the variables $w_n^{\lr{+}}$ and $w_n^{\lr{-}}$ which are proportional to the probabilities to find the elements $\lrc{n, +}$ or $\lrc{n, -}$ at the top of the stack (provided the stack contains more than one element). Then $w_n = w_n^{\lr{+}} - w_n^{\lr{-}}$. We thus arrive at the following random process for stochastic evaluation of $w_n^{\lr{\pm}}$. At each discrete time step one performs at random one of the following actions:
\begin{itemize}
  \item With probability $\mathcal{N}^{-1}$ add new element $\lrc{1, +}$ to the stack.
  \item With probability $2 \, c^{-1}$ increase the topmost element in the stack by $1$ and do not change its sign.
  \item With probability $|\lambda| \, c$ decrease the topmost element in the stack by $1$ (if it is greater than one) and multiply its sign by the sign of $\lambda$.
  \item With probability $\mathcal{N}\, c^{-2}$ pop the two elements $\lrc{n, s_1}$ and $\lrc{m, s_2}$ from the stack (provided there are more than two elements) and push the element $\lrc{n + m + 1, s_1 s_2}$ to the stack.
  \item Otherwise empty the stack and push into it a single element $\lrc{1, +}$.
\end{itemize}
Note that for positive $\lambda$ elements with the minus sign are not generated, so that $w_n^{\lr{-}} \equiv 0$ and the random process automatically reduces to the one described in Section \ref{sec:recursive_process}.

 The inequalities (\ref{probability_ineq}) read now:
\begin{eqnarray}
\label{brezin_process_inequalities}
  \mathcal{N}^{-1} + 2 \, c^{-1} + |\lambda| c + \mathcal{N}\, c^{-2} \le 1,
  \quad \mathcal{N} > 0, \quad c > 0
\end{eqnarray}
As discussed in Subsection \ref{subsec:phi4_general}, in order to increase the efficiency of the algorithm it is advantageous to saturate this inequality. It is easy to see that at the same time we saturate the upper bound on the absolute value of the coupling constant $\lambda$. Maximizing this upper bound with respect to $\mathcal{N}$ and $c$, we see that $|\lambda|$ cannot exceed the value $\bar{\lambda} = 1/16 = 3/4 \, \lambda_c$, where $\lambda_c = 1/12$ is the convergence radius of the planar perturbative expansion which can be found from the exact solution of the matrix model (\ref{matrix_model_def}) \cite{Brezin:78:1}. Thus the described random process covers only some finite subrange of coupling constants for which the model (\ref{matrix_model_def}) is defined. It is easy to understand the origin of this limitation: in fact the random process described above simulates an ensemble of diagrams with an arbitrary number of external legs, with the weight of each diagram being proportional to $\lambda^{N_v}$, where $N_v$ is the number of vertices. The number of open diagrams with a given number of vertices is obviously larger than the number of closed diagrams, hence the sums over open diagrams have smaller convergence radius.

\begin{figure}[h]
  \includegraphics[width=6cm, angle=-90]{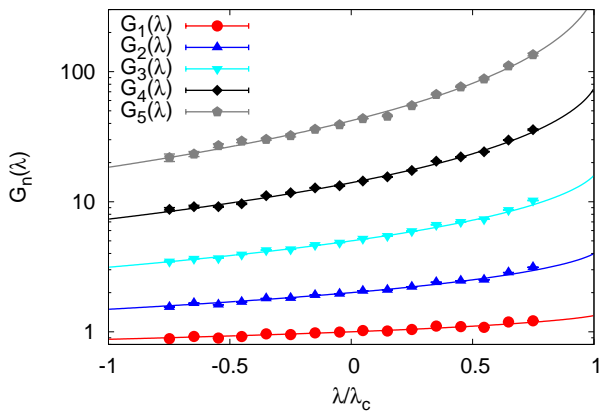}\\
  \caption{Green functions $G_n\lr{\lambda}$ in (\ref{matrix_model_SDs}) versus the coupling constant $\lambda$ for $n = 1, \ldots, 5$, obtained after $N = 10^6$ discrete time steps of the described algorithm at fixed $c = 8$ and with $\mathcal{N}$ given by ``Branch 1'' of (\ref{nn_vs_lambda}). The error bars are smaller than the symbols on the plot. Exact results of \cite{Brezin:78:1} are plotted with solid lines. }
  \label{fig:brezin_moments}
\end{figure}

 For $|\lambda| < \bar{\lambda}$, there is a continuous set of $\mathcal{N}$, $c$ which saturate the inequality (\ref{brezin_process_inequalities}). One can, for example, fix $c$ and express $\mathcal{N}$ as a function of $\lambda$:
\begin{eqnarray}
\label{nn_vs_lambda}
\mathcal{N} = \frac{c^2 - 2 c - |\lambda| c^3 \pm \sqrt{c^3\lr{c |\lambda| - 1}\lr{4 - c + c^2 |\lambda|}}}{2}  .
\end{eqnarray}
We call the solution with the minus sign in front of the square root ``Branch~1'' and the other solution ``Branch~2''.

\begin{figure}[h]
  \includegraphics[width=6cm, angle=-90]{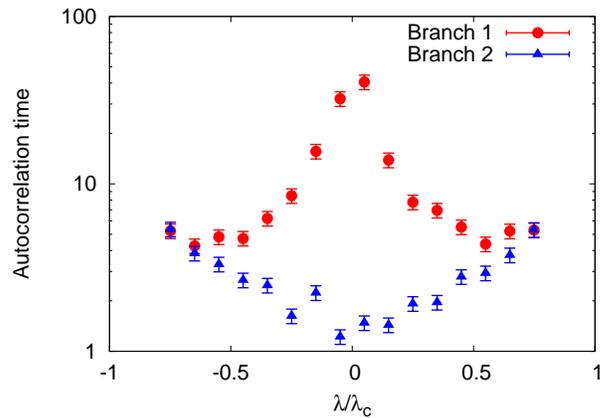}\\
  \caption{Autocorrelation time of the random process described in Subsection \ref{subsec:matrix_model} as the function of the coupling constant $\lambda$ at fixed $c = 8$ and for different choices of $\mathcal{N}$ in (\ref{nn_vs_lambda}).}
 \label{fig:autocorrelation_time}
\end{figure}

 On Fig. \ref{fig:brezin_moments} we plot the Green functions $G_n\lr{\lambda}$ evaluated using the described random process as functions of $\lambda$ up to $n=5$. These results were obtained after $N = 10^6$ discrete time steps at fixed $c = 8$ and with $\mathcal{N}$ given by the ``Branch 1'' of (\ref{nn_vs_lambda}). The error bars are smaller than the symbols on the plot. Solid lines are the exact results for $G_n\lr{\lambda}$ in the planar approximation, obtained using the saddle point method \cite{Brezin:78:1}.

\begin{figure}[h]
  \includegraphics[width=6cm, angle=-90]{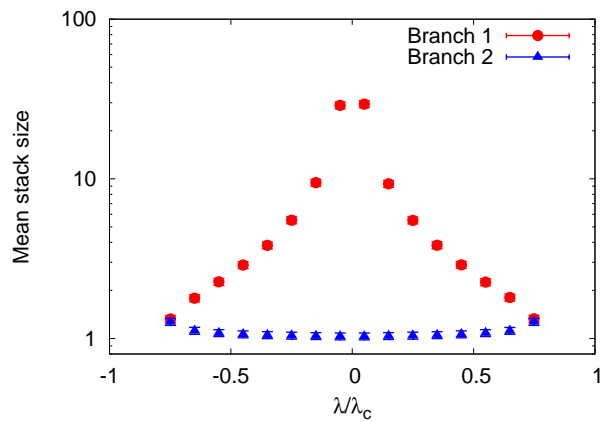}\\
  \caption{Mean stack size of the random process described in Subsection \ref{subsec:matrix_model} as the function of the coupling constant $\lambda$ at fixed $c = 8$ and for different choices of $\mathcal{N}$ in (\ref{nn_vs_lambda}).}
  \label{fig:mean_stack_size}
\end{figure}

 Autocorrelation time and mean stack size for the described random process are plotted on Figs. \ref{fig:autocorrelation_time} and \ref{fig:mean_stack_size}, respectively, as the functions of the coupling constant $\lambda$. The observable used to define the autocorrelation time was the sum of all numbers in the stack. First, we note that ``Branch 1'' is more advantageous for simulations, since with larger mean stack size one can gain more statistics. However, in this case the autocorrelation time is also larger. Interestingly, for this branch both the autocorrelation time and the mean stack size have maximum near $\lambda = 0$ rather than near the ``critical point'' of the random process $\bar{\lambda}$. For ``Branch 2'', these quantities increase slowly towards $\bar{\lambda}$.

\begin{figure}[h]
  \includegraphics[width=6cm, angle=-90]{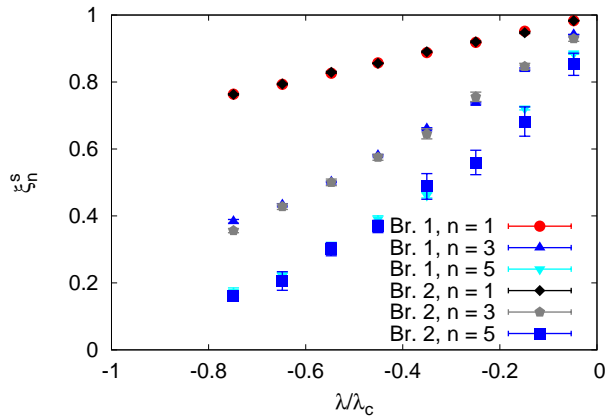}\\
  \caption{The quantity $\xi^s_n$ in (\ref{matrix_model_xisn_def}) for $n = 1, 3, 5$, as the function of the coupling constant $\lambda$ at fixed $c = 8$ and for different choices of $\mathcal{N}$ in (\ref{nn_vs_lambda}). ``Br. 1,2'' is for ``Branch 1,2''. }
  \label{fig:sign_problem_severity}
\end{figure}

 In order to characterize the strength of the sign problem, we consider the quantity
\begin{eqnarray}
\label{matrix_model_xisn_def}
 \xi^{s}_n = \lr{w_n^{\lr{+}} - w_n^{\lr{-}}}/\lr{w_n^{\lr{+}} + w_n^{\lr{-}}} .
\end{eqnarray}
$\xi^{s}_n = 1$ if the random process generates only elements with the plus sign and $\xi^s_n = 0$ if the numbers of pluses and minuses exactly cancel. In practice, it is advantageous to have as large $\xi^{s}_n$ as possible, so that the difference $w_n^{\lr{+}} - w_n^{\lr{-}}$ can be estimated with maximal precision.  $\xi^{s}_n$ are plotted on Fig. \ref{fig:sign_problem_severity} as the function of $\lambda$ for $n = 1, 3, 5$. For $\lambda < 0$, $\xi_s\lr{\lambda}$ decreases with $\lambda$ and $n$. The sign cancelation is thus moderate for $n=1$ ($\xi^s_1\lr{-\bar{\lambda}} \approx 0.75$) and becomes more and more important for higher-order correlators - $\xi^s_5$ at $\lambda = - \bar{\lambda}$ is close to zero. It is interesting that $\xi^{s}_n$ are almost equal for two different choices of $\mathcal{N}$ in (\ref{nn_vs_lambda}).

 Thus there are no indications of severe critical slowing down in the whole range of possible coupling constants $-\bar{\lambda} < \lambda < \bar{\lambda}$. The sign problem is also moderate for low-order correlators, but becomes more severe for higher-order correlators. It could be extremely interesting to extend the applicability of the described random process up to $\lambda = \lambda_c$ while preserving these attractive features of the algorithm.

\subsection{Random planar surfaces: the Weingarten model}
\label{subsec:weingarten}

 Weingarten model \cite{Weingarten:80:1, Eguchi:82:3} is a lattice field theory which in the large-$N$ limit reproduces the sum over all closed surfaces with genus one on the hypercubic lattice. The action for each surface is proportional to its area, thus the model can be considered as a lattice regularization of bosonic strings with Nambu-Goto action. Although this model does not have a nontrivial continuum limit for any space dimensionality \cite{Durhuus:84:1}, the structure of the functional integral and of the Schwinger-Dyson equations in this model are similar to those in large-$N$ non-Abelian lattice gauge theory, and the analysis of this model might be helpful for the extension of the approach described here to non-Abelian gauge theories. In order to derive the Schwinger-Dyson equations, it is convenient to consider the reduced Weingarten model \cite{Eguchi:82:3}, which in the large-N limit is equivalent to the original model, similarly to the Eguchi-Kawai model for non-Abelian lattice gauge theory. It can be shown that in contrast to reduced lattice gauge theories, for the reduced Weingarten model additional twisting is not necessary \cite{Kawai:83:1}.

\begin{figure}
  \includegraphics[width=6cm]{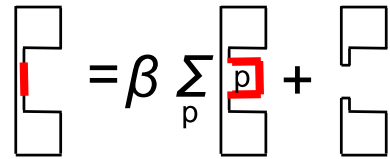}\\
  \caption{Schematic illustration of the structure of the loop equations (\ref{weingarten_loop_equations}) in the Weingarten model (\ref{weingarten_model_def}). These equations should hold for any link (marked by a thick line) which belongs to the loop.}
  \label{fig:weingarten_loop_equations}
\end{figure}

 The reduced model is defined by an integral over complex $N \times N$ matrices $U_{\mu} \equiv U^{\dag}_{-\mu}$ with $\mu = 1, \ldots, D$:
\begin{widetext}
\begin{eqnarray}
\label{weingarten_model_def}
\mathcal{Z}\lr{\beta} = \int \mathcal{D} U_{\mu} \,
\expa{ -N \sum \limits_{\mu = 1}^{D} \tr\lr{U_{\mu} U^{\dag}_{\mu}}
+ N \beta \sum \limits_{\mu \neq \nu = 1}^{D} \tr\lr{U_{\mu} U_{\nu} U_{\mu}^{\dag} U_{\nu}^{\dag}} }  .
\end{eqnarray}
\end{widetext}
 If one treats the second term in the exponent in (\ref{weingarten_model_def}) as a perturbation and expands $Z\lr{\beta}$ in powers of $\beta$, the resulting sum over planar diagrams is equivalent to the sum over all possible closed surfaces of genus one on the lattice with the weight $\beta^{|S|}$, where $|S|$ is the area of each surface.

 A basic observable in this model is the sum over all planar surfaces which are bounded by some closed loop $C$. The loop $C$ can be uniquely specified by a sequence $\lrc{\mu_1, \ldots, \mu_n}$, where $\mu$'s take the values $\pm 1, \ldots, \pm D$. In order to reconstruct the loop $C$ from the sequence, one should start from an arbitrary point on a hypercubical lattice and move along one link in the direction $\mu_1$, forward if $\mu_1$ is positive and backward if $\mu_1$ is negative. From this new position one should similarly move in the direction $\mu_2$, and so on. From the diagrammatic expansion one can see that such a sum over surfaces is given by the following correlator:
\begin{eqnarray}
\label{weingarten_wilson_loop}
 W\lr{C} = W\lr{\mu_1, \ldots, \mu_n} =
 \vev{ \frac{1}{N}\, \tr\lr{U_{\mu_1} \ldots U_{\mu_n}} },
\end{eqnarray}
where one takes the conjugate variable $U_{|\mu_A|}^{\dag}$ if $\mu_A$ is negative. This observable is similar to Wilson loop in lattice gauge theory, but, unlike the Wilson loop, it does not have a ``zigzag symmetry'' \cite{Polyakov:99:1}: passing a link forward and immediately backward changes the value of the Wilson loop. In the large-$N$ limit the single-loop observables factorize, which allows us to obtain a closed set of Schwinger-Dyson equations for $W\lr{\mu_1, \ldots, \mu_n}$ \cite{Weingarten:80:1, Eguchi:82:3}:
\begin{widetext}
\begin{eqnarray}
\label{weingarten_loop_equations}
W\lr{\mu_1, \mu_2} = \delta\lr{\mu_1, -\mu_2} + \beta \sum \limits_{|\mu| \neq |\mu_1|} W\lr{\mu, \mu_1, -\mu, \mu_2}
\nonumber \\
W\lr{\mu_1, \ldots, \mu_n}
 =
\delta\lr{\mu_1, -\mu_2} W\lr{\mu_3, \ldots, \mu_n}
 +
\delta\lr{\mu_1, -\mu_n} W\lr{\mu_2, \ldots, \mu_{n-1}}
 + \nonumber \\ +
\sum \limits_{A = 3}^{n-1} W\lr{\mu_2, \ldots, \mu_{A-1}} \, W\lr{\mu_{A+1}, \ldots, \mu_n} \delta\lr{\mu_1, -\mu_A}
 + \nonumber \\ +
\beta \sum \limits_{|\mu| \neq |\mu_1|} W\lr{\mu, \mu_1, -\mu, \mu_2, \ldots, \mu_n}, \quad n > 2.
\end{eqnarray}
These equations should hold for any lattice link $\mu_k$ belonging to the loop $C$, but the resulting system of equations is redundant, and it is sufficient to consider only one link $\mu_1$ on the loop. The equations (\ref{weingarten_loop_equations}) are schematically illustrated on Fig. \ref{fig:weingarten_loop_equations}, where the link $\mu_1$ is marked by a thick line.

 We see that the equations (\ref{weingarten_loop_equations}) again take the form similar to (\ref{random_process_eq}). Let us now define the ``renormalized'' observable $w\lr{\mu_1, \ldots, \mu_n}$ by rescaling $W\lr{\mu_1, \ldots, \mu_n}$ by the factors $\mathcal{N}$ and $q$ as $W\lr{\mu_1, \ldots, \mu_n} = \mathcal{N} q^{n} w\lr{\mu_1, \ldots, \mu_n}$. One can interpret the factor $q^{n}$ as the mass attached to the boundaries of random surfaces, somewhat like the bare quark mass in QCD. The equations (\ref{weingarten_loop_equations}) then take the following form:
\begin{eqnarray}
\label{weingarten_loop_equations_renorm}
w\lr{\mu_1, \mu_2} = \lr{\mathcal{N} q^2}^{-1} \delta\lr{\mu_1, -\mu_2} + \beta q^2 \sum \limits_{|\mu| \neq |\mu_1|} w\lr{\mu, \mu_1, -\mu, \mu_2}
\nonumber \\
w\lr{\mu_1, \ldots, \mu_n}
 =
q^{-2} \delta\lr{\mu_1, -\mu_2} w\lr{\mu_3, \ldots, \mu_n}
 +
q^{-2} \delta\lr{\mu_1, -\mu_n} w\lr{\mu_2, \ldots, \mu_{n-1}}
 + \nonumber \\ +
\mathcal{N} q^{-2} \, \sum \limits_{A = 3}^{n-1} w\lr{\mu_2, \ldots, \mu_{A-1}} \, w\lr{\mu_{A+1}, \ldots, \mu_n} \delta\lr{\mu_1, -\mu_A}
 + \nonumber \\ +
\beta q^2 \, \sum \limits_{|\mu| \neq |\mu_1|} w\lr{\mu, \mu_1, -\mu, \mu_2, \ldots, \mu_n}, \quad n > 2.
\end{eqnarray}
\end{widetext}

 Let us now devise a random process of the type described in Section \ref{sec:recursive_process}, which solves stochastically these equations. The configuration space is now a stack which contains closed loops, that is, sequences of indices $\mu = \pm 1, \ldots, \pm D$. The desired random process is defined by the following possible actions at each discrete time step:
\begin{description}
  \item[Create a new loop:] With probability $2 D \mathcal{N}^{-1} q^{-2}$ create a new elementary loop $C = \lrc{\mu, -\mu}$, where $\mu = \pm 1, \ldots, \pm D$ is random (either positive or negative).
  \item[Join loops:] With probability $2 D \mathcal{N} q^{-2}$ pop the two loops $C_1 = \lrc{\mu_1, \ldots, \mu_n}$, $C_2 = \lrc{\nu_1, \ldots, \nu_m}$ from the stack and form a new loop $C$ by joining the loops $C_1$, $C_2$ with a link in the random direction $\mu$ (either positive or negative): $C = \lrc{\mu_1, \ldots, \mu_n, \mu, \nu_1, \ldots, \nu_m, -\mu}$. This action can only be performed if there are more than two loops in the stack.
  \item[Flatten loop:] If the three links in the end of the sequence on the top of the stack form a boundary of the plaquette, that is, if the topmost loop has the form $C = \lrc{\mu_1, \ldots, \mu_n, \mu, \nu, -\mu}$ for some $\mu$ and $\nu$, replace these three links by a single link in the direction $\nu$ with probability $\beta q^2$: $C' = \lrc{\mu_1, \ldots, \mu_n, \nu}$.
  \item[Append to loop:] With probability $4 D q^{-2}$ append a pair $\lrc{\mu, -\mu}$, where $\mu$ is random (either positive or negative), to the topmost sequence $\lrc{\mu_1, \ldots, \mu_n}$ in the stack as $\lrc{\mu, -\mu, \mu_1, \ldots, \mu_n}$ or $\lrc{\mu, \mu_1, \ldots, \mu_n, -\mu}$. The probabilities of these two choices are equal.
  \item[Restart:] Otherwise start with a stack containing an elementary random loop $C = \lrc{\mu, -\mu}$, where $\mu = \pm 1, \ldots, \pm D$ is chosen randomly.
\end{description}

 Again assuming that the sum of the probabilities of all possible actions is equal to one and the probability of ``Restart'' events is minimized, we obtain an equation relating $\beta$, $\mathcal{N}$ and $q$:
\begin{eqnarray}
\label{weingarten_param_relation}
 \beta q^2 + 2 D q^{-2} \lr{\mathcal{N} + \mathcal{N}^{-1} + 2} = 1
\end{eqnarray}
Maximization with respect to $\mathcal{N}$ yields the relation between $q$ and $\beta$:
\begin{eqnarray}
\label{weingarten_q_vs_beta}
q = \sqrt{\frac{1 \pm \sqrt{1 - 32 D \beta}}{2 \beta}}  .
\end{eqnarray}
We call the solution with the minus sign in front of the square root ``Branch 1'' and the other solution ``Branch 2''.

\begin{figure}
  \includegraphics[width=6cm, angle=-90]{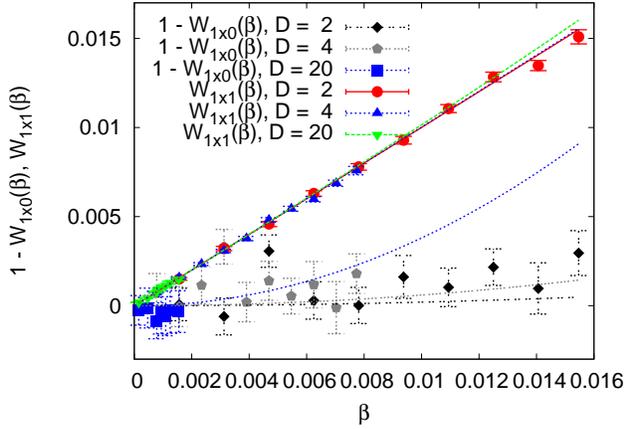}\\
  \caption{The observables $W_{1\times0} \equiv W\lr{\mu, -\mu}$ ($1 \times 0$ loop) and $W_{1 \times 1} \equiv W\lr{\mu, \nu, -\mu, -\nu}$ ($1 \times 1$ loop) in the Weingarten model as a functions of the coupling constant $\beta$ for different dimensions $D$. Solid line corresponds to the first two terms in the perturbative expansion: $W\lr{\mu, -\mu} = 1 + 2 \lr{D - 1} \beta^2 + O\lr{\beta^4}$, $W\lr{\mu, \nu, -\mu, -\nu} = \beta + 8 \lr{D - 1} \beta^3 + O\lr{\beta^5}$. The data were obtained after $10^7$ iterations of the algorithm described above.}
  \label{fig:wgtn_w1x1_w1x0}
\end{figure}

 The value of $\beta$ in (\ref{weingarten_q_vs_beta}) cannot exceed the critical value $\bar{\beta}\lr{D} = 1/\lr{32 D}$. As we have already seen on the example of the matrix model, this critical value does not necessarily coincide with the true critical point $\beta_c\lr{D}$ at which the sum over planar surfaces diverges. Indeed, $\bar{\beta}\lr{D}$ does not exceed the lower bound $\beta_c\lr{D} > \lr{24 \lr{D - 1}}^{-1}$ obtained in \cite{Eguchi:82:3}, and is significantly lower than the critical values obtained numerically in \cite{Kawai:83:1}. In fact, for $\beta = \bar{\beta}_{D}$ all the observables are still dominated by the lowest-order perturbative contributions. Expectation values of the observables $W_{1\times0} \equiv W\lr{\mu, -\mu}$ ($1 \times 0$ loop) and $W_{1 \times 1} \equiv W\lr{\mu, \nu, -\mu, -\nu}$ ($1 \times 1$ loop), which were obtained after $10^7$ iterations of the random process described above (with $q$ given by ``Branch 1'' of (\ref{weingarten_q_vs_beta})), are plotted on Fig. \ref{fig:wgtn_w1x1_w1x0} as the functions of the coupling constant $\beta$. Solid line corresponds to the first two terms in the perturbative expansion: $W_{1 \times 0} = 1 + 2 \lr{D - 1} \beta^2 + O\lr{\beta^4}$, $W_{1 \times 1} = \beta + 8 \lr{D - 1} \beta^3 + O\lr{\beta^5}$. Within statistical errors, one sees only the lowest-order perturbative contributions. It should be stressed that the proposed random process implements stochastic summation of diagrams of \emph{all} orders, but due to the smallness of $\beta < \bar{\beta}\lr{D}$, a very large computational time is required to see the contributions of higher-order terms.

\begin{figure}
  \includegraphics[width=4cm]{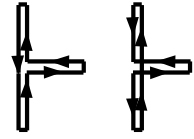}\\
  \vspace{-0.5cm}
  \caption{Two simple configurations of branched polymers on the lattice. The configurations on the left and on the right count as different configurations.}
  \label{fig:branched_polymers_illustrated}
\end{figure}

 Note that when the coupling constant $\beta$ tends to zero (that is, the ``bare string tension'' of the random surfaces tends to infinity) and $q$ lies between the two solutions of (\ref{weingarten_q_vs_beta}), the above random process describes just the growth of ``branched polymers'', whose branches are bosonic random walks and hence correspond to particles rather than ``strings''.  These branches consist of loops in which every lattice link is passed twice and which hence sweep out zero area.

 Taking the limit $\beta \rightarrow 0$ in (\ref{weingarten_q_vs_beta}), we find that the minimal value of $q$ is $\bar{q} = \sqrt{8 D}$. In order to understand this critical value we first note that in the limit $\beta \rightarrow 0$ the observables $W\lr{\mu_1, \ldots, \mu_n}$ are all equal to one if the links $\mu_1, \ldots, \mu_n$ form a loop which sweeps out zero area and zero otherwise. The probabilities to encounter such loops in the described random process is hence proportional to $w\lr{\mu_1, \ldots, \mu_n} \sim q^{-n}$. Simple examples of such loops, which can be also thought of as the random tree-like graphs on the lattice, are shown on Fig. \ref{fig:branched_polymers_illustrated}. Now imagine adding to some loop $k$ links stemming from some lattice site. Since the loop includes each link twice, the probability decreases by $q^{-2 k}$. The number of possible configurations of $k$ links is $\lr{2 \times 2 D}^{k}$ since each of $k$ links can point along any of $D$ directions both forward and backward. An additional factor of $2$ appears since the zero-area loops pass twice through each point and the new links can be inserted between the links pointing either forward or backward (for example, compare the configurations on the left and on the right of Fig. \ref{fig:branched_polymers_illustrated}). Finally, one can add any number $k = 1, 2, \ldots, \infty$ of branches to any point belonging to the branched polymer. At the criticality, adding any number of random links to some configuration should not change its overall weight. Therefore, the change of the weight due to the added links times the number of ways to add them should be equal to unity. We are thus led to the following equation for $\bar{q}$:
\begin{eqnarray}
\label{branched_polymers_weight}
\sum \limits_{k = 1}^{+\infty} \lr{4 D q^{-2}}^k = \frac{4 D q^{-2}}{1 - 4 D q^{-2}} = 1  ,
\end{eqnarray}
or $8 D \bar{q}^{- 2} = 1$. Thus, in the limit $\beta \rightarrow 0$ we indeed reproduce branched polymers with the correct critical behavior.

\begin{figure}
  \includegraphics[width=6cm, angle=-90]{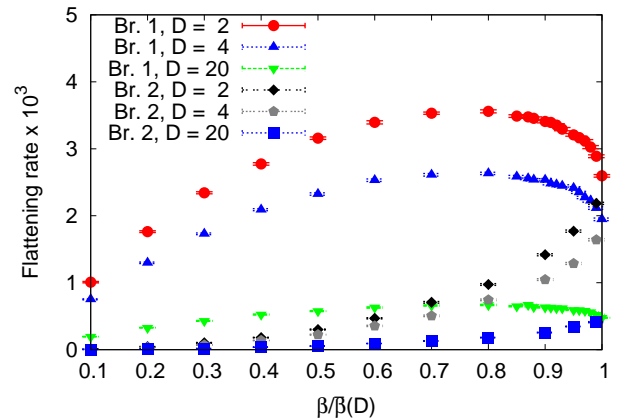}\\
  \caption{Mean rate of ``flattening'' events for the random process solving the loop equations in the Weingarten model as a function of the coupling constant $\beta$ at different dimensions $D$ and for different choices of $q$ in (\ref{weingarten_q_vs_beta}). ``Br. 1,2'' is for ``Branch 1, 2''.}
  \label{fig:wgtn_mfr}
\end{figure}

\begin{figure*}
  \includegraphics[width=6cm, angle=-90]{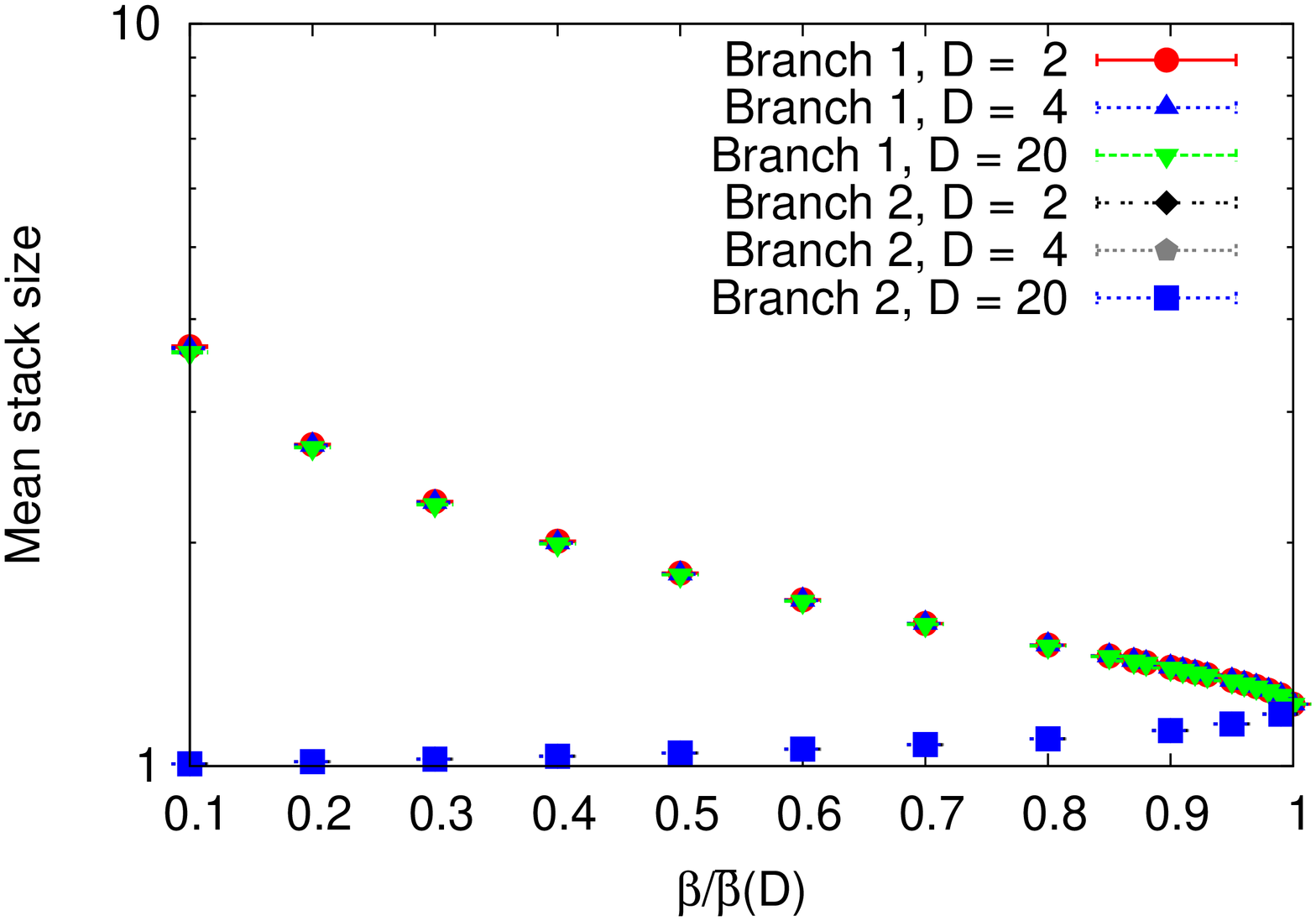}
  \includegraphics[width=6cm, angle=-90]{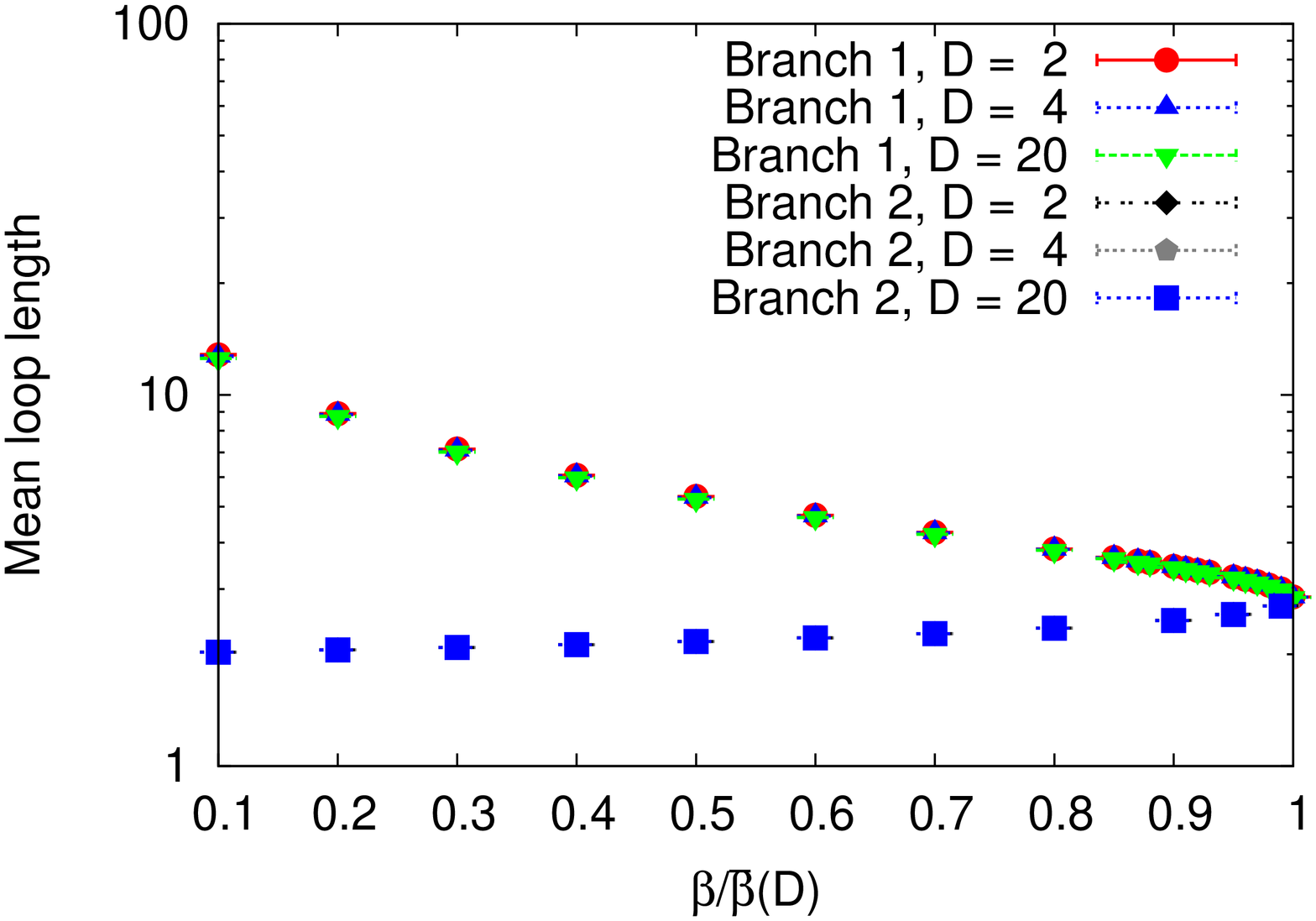}\\
  \caption{Mean stack size (on the left) and mean length of the topmost loop in the stack (on the right) for the random process solving the loop equations in the Weingarten model as a function of the coupling constant $\beta$ at different dimensions $D$ and for different choices of $q$ in (\ref{weingarten_q_vs_beta}).}
  \label{fig:wgtn_mss_mll}
\end{figure*}

 At nonzero $\beta$, deviation from trivial branched polymer configurations can be characterized by the rate of the ``Flatten loop'' events. Indeed, since the probability of $n$ such events is proportional to $\beta^n$, such sequence of events corresponds to a random surface (which is in general open) consisting of $n$ lattice plaquettes, plus some number of random trees. One can therefore think of the described random process as of the process of drawing random loops which sweep out random planar surfaces. The average rate of ``flattening'' events is plotted on Fig. \ref{fig:wgtn_mfr} on the right as a function of the coupling constant $\beta$ for different dimensions $D$ and for different choices of $q$ in (\ref{weingarten_q_vs_beta}). One can see that in the whole range of coupling constants $0 < \beta < \bar{\beta}\lr{D}$ the rate of flattening events is numerically very small. On the other hand, the number of links in the loops, as well as the number of loops stored in the stack, are quite large. Mean stack size and mean length of the topmost loop in the stack are plotted on Fig. \ref{fig:wgtn_mss_mll} as a function of the coupling constant $\beta$ for different dimensions $D$ and for different choices of $q$ in (\ref{weingarten_q_vs_beta}). One can conclude therefore that the ``branched polymers'' actually dominate in the properties of the described random process. Critical behavior of these random trees is universal for any dimension $D$, that is why such observables as the mean stack size or the mean loop length, which are mainly sensitive to the length of loops rather than to the area of random surfaces, practically do not depend on space dimensionality.

 While the closed planar surfaces in the vicinity of the true critical point $\beta_c\lr{D}$ of the Weingarten model are also dominated by ``branched polymers'' \cite{Durhuus:84:1}, in our ensemble of open random surfaces this dominance can be thought of as the manifestation of the tachyonic instability of open, rather than closed, strings. The fact that the critical coupling $\bar{\beta}\lr{D}$ in our case is smaller than the true critical point $\beta_c\lr{D}$ can be explained by the fact that the number of open surfaces with a given area is obviously larger than the number of closed surfaces with the same area. The true critical coupling constant $\beta_c\lr{D}$ can be quite easily found by a very simple re-weighting procedure, which will be described in details in a separate publication.

\section{Resummation of divergent series and random processes with memory}
\label{sec:rps_with_mem}

 In Section \ref{sec:SDs_stochastic_solution} we have described a stochastic method for the solution of the Schwinger-Dyson equations for theories with noncompact variables. This method works only at small coupling constants and implements stochastic summation of perturbative series. For theories with compact variables, such as nonlinear $\sigma$-models or lattice gauge theories, the structure of the Schwinger-Dyson equations is such that the described method can be straightforwardly applied only in the strong coupling regime, where one can stochastically sum all terms in the strong-coupling expansion. However, the continuum limit of such theories typically corresponds to the weak-coupling limit. An additional complication is that for physically interesting theories the observables cannot be expressed as convergent power series in the small coupling constant $g$, but rather contain non-analytic part which is typically of the form $\expa{-c/g^2}$ with some constant $c$ \cite{Parisi:77:1}.

 In this Section we point out one possible way to deal with this problem. The basic idea is to absorb non-perturbative corrections into some self-consistent redefinition of the expansion parameter \cite{Parisi:77:1, Kazakov:94:1}. Recently, a similar resummation method was also considered in \cite{Prokofev:10:2}. Solving the self-consistency condition leads to the concept of a nonlinear random process with memory \cite{Frank:04:1}, in which all previous history of the process is used to estimate the value of the self-consistent expansion parameter.

 Let us illustrate this idea on the simplest example of $O\lr{N}$ sigma model in the limit of large $N$. The model is defined by the following path integral over unit $N$-component vectors $n\lr{x}$ living on the sites of the $D$-dimensional hypercubic lattice:
\begin{eqnarray}
\label{sigma_model_pf}
 \mathcal{Z} = \int\limits_{|n\lr{x}|=1} \mathcal{D}n(x) \expa{ \frac{N}{\lambda} \, \sum \limits_{<xy>} n\lr{x} \cdot n\lr{y} },
\end{eqnarray}
where summation goes over all neighboring lattice sites. Despite its simplicity, this model in $D = 2$ dimensions is asymptotically free and has a mass gap which depends nonperturbatively on the coupling constant $\lambda$. Schwinger-Dyson equations in this theory can be written in terms of the two-point function $\xi\lr{x, y} = \vev{n\lr{x}~\cdot~n\lr{y}}$, $\xi\lr{x} \equiv \xi\lr{x, 0}$ as:
\begin{eqnarray}
\label{sigma_model_sd}
 \xi\lr{x} = \frac{1}{\lambda} \sum \limits_{\mu} \lr{\xi\lr{x \pm e_{\mu}} - \xi\lr{x} \xi\lr{\pm e_{\mu}}} + \delta\lr{x, 0}  .
\end{eqnarray}
Clearly, these equations have the structure similar to (\ref{random_process_eq}), but the inequalities (\ref{probability_ineq}) are satisfied only for sufficiently large $\lambda$, that is, in the strong-coupling regime. Therefore, the continuum limit at $\lambda \rightarrow 0$ cannot be reached by the method described in Section \ref{sec:SDs_stochastic_solution}.

 Let us, however, rewrite the equation (\ref{sigma_model_sd}) as
\begin{eqnarray}
\label{sigma_model_sd_as_rw}
 \xi\lr{x} = \frac{1}{\lambda + \sum \limits_{\mu} \xi\lr{\pm e_\mu}} \, \lr{ \sum \limits_{\mu} \xi\lr{x \pm e_{\mu}} + \lambda \delta\lr{x} } ,
\end{eqnarray}
and introduce the ``hopping parameter''
\begin{eqnarray}
\label{sigma_model_kappa_def}
\kappa = \frac{1}{\lambda + \sum \limits_{\mu} \xi\lr{\pm e_\mu}}  .
\end{eqnarray}
Now the equation (\ref{sigma_model_sd}) in the form (\ref{sigma_model_sd_as_rw}) looks like the equation for the free massive scalar propagator on the lattice with the mass $m^2 = \kappa^{-1} - 2 D$ in lattice units. Note that in the weak-coupling limit $\lambda \rightarrow 0$ $\xi\lr{\pm e_\mu} \rightarrow 1$, $m^2 \rightarrow 0$ and we approach the continuum limit.

 Let us now solve the equation (\ref{sigma_model_sd_as_rw}) stochastically, assuming that $\xi\lr{x}$ is proportional to the stationary probability distribution $w\lr{x}$ of some random process: $\xi\lr{x} = c \, w\lr{x}$, $\sum \limits_{x} w\lr{x} = 1$. From (\ref{sigma_model_sd_as_rw}) we get $c = \frac{\lambda \kappa}{1 - 2 D \kappa}$. The equation (\ref{sigma_model_sd_as_rw}) now looks as
\begin{eqnarray}
\label{sigma_model_rw_eq}
w\lr{x} = \kappa \sum \limits_{\mu} w\lr{x \pm e_\mu} + \lr{1 - 2 D \kappa} \, \delta\lr{x} .
\end{eqnarray}
Combining this equation with the definition (\ref{sigma_model_kappa_def}), it is easy to show that $\kappa$ obeys the following self-consistency condition:
\begin{eqnarray}
\label{sigma_model_kappa_return_prob}
 \kappa = \frac{1}{2 D + \lambda w\lr{0}}  .
\end{eqnarray}
The equation (\ref{sigma_model_rw_eq}) has the form (\ref{random_process_eq}) without the nonlinear term and thus can be interpreted as the equation for the stationary probability distribution of the position of an ordinary bosonic random walk, defined by the following possible actions at each discrete time step:
\begin{description}
  \item[Move:] With probability $2 D \kappa$ move along the random unit lattice vector $\pm e_{\mu}$.
  \item[Restart:] With probability $\lr{1 - 2 D \kappa}$ start again at the origin $x = 0$.
\end{description}
This ensures that $w\lr{0} > 0$ and hence $\kappa$ never exceeds its critical value $\kappa_c = \lr{2 D}^{-1}$. Therefore $\xi\lr{x}$ and $w\lr{x}$ can be expanded in powers of $\kappa$.

 Thus we have defined a new expansion parameter $\kappa$, which should obey the self-consistency equation (\ref{sigma_model_kappa_return_prob}), and obtained a well-defined convergent expansion, namely, the sum over all paths on the lattice with the weight $\kappa^L$, where $L$ is the length of the path. Note that the quantity $\xi\lr{\pm e_\mu}$ in fact plays the role similar to the gluon condensate (which is expressed in terms of the mean plaquette in lattice theory) in non-Abelian gauge theory: one can absorb all the divergences into the self-consistent definition of condensates \cite{Parisi:77:1, Kazakov:94:1}.

 The final step in the construction of the nonlinear random process which solves the equations (\ref{sigma_model_sd}) is the solution of the self-consistency equation (\ref{sigma_model_kappa_return_prob}). One possible solution is to use the iterations
\begin{eqnarray}
\label{sigma_model_iterations}
\kappa_{i+1} = \frac{1}{2 D + \lambda w\lr{0; \kappa_i}}  .
\end{eqnarray}
Here $w\lr{0; \kappa}$ is the return probability of a bosonic random walk with hopping parameter $\kappa$. In practice, one should simulate the bosonic random walk at fixed $\kappa = \kappa_i$ for some number $T$ of discrete time steps, and then estimate $w\lr{0; \kappa_i}$ as $w\lr{0; \kappa_i} \approx t\lr{0}/T$, where $t\lr{0}$ is the number of discrete time steps spent at $x = 0$. From (\ref{sigma_model_iterations}) one then gets $\kappa_{i+1}$, and the process is repeated until the value of $\kappa$ stabilizes with sufficient numerical precision. We call such algorithm ``Algorithm A''. One can also consider an ultimate case, for which the return probability is updated and estimated as $t\lr{0}/t$ every time the point $x = 0$ is reached. Now $t$ is the time from the start of the random process and $t\lr{0}$ is the number of time steps spent at $x = 0$. Such algorithm will be called ``Algorithm B''.

 Mathematically, such random processes are not Markov processes, since the transition probabilities at each next step depend (via $\kappa_i$) on the behavior of the process at all previous time steps. Stationary probability distributions of such processes obey nonlinear equations (such as (\ref{sigma_model_sd})) \cite{Frank:04:1}, and and hence they are also called nonlinear random processes.

 As an interesting side remark, let us discuss such a theory at finite temperature, which is described by a bosonic random walk on the cylinder. Clearly, an ordinary bosonic random walk does not feel this compactification of space, and its stationary probability distribution is just a periodic linear combination of the corresponding distribution in infinite space. Such behavior cannot lead to any nonlinear finite-temperature effects such as phase transitions. On the other hand, if the parameters of the random walk depend on the return probability, as in (\ref{sigma_model_kappa_return_prob}), there is a nonlinear feedback mechanism since in the compactified space the returns are more likely. Thus finite temperature indeed affects the local behavior of the random walker with memory and might lead to interesting critical phenomena.

\begin{figure}
  \includegraphics[width=6cm, angle=-90]{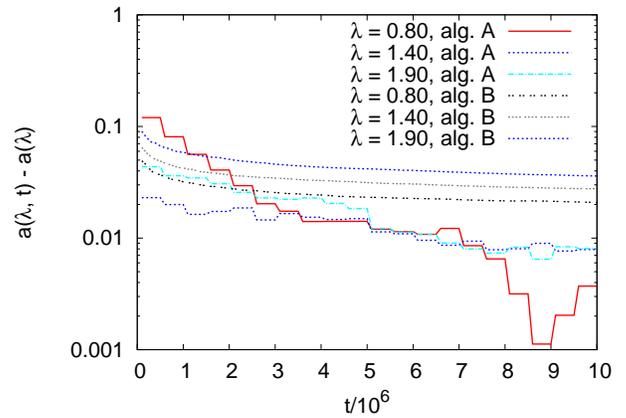}\\
  \caption{The process of convergence of the random process with memory which solves the Schwinger-Dyson equations (\ref{sigma_model_sd}) for $D = 2$. The quantity plotted is the estimate of the lattice spacing $a\lr{\lambda, t}$ after $t$ discrete time steps, with the exact result $a\lr{\lambda, t \rightarrow \infty}$ subtracted.}
  \label{fig:sigma_model_convergence}
\end{figure}

\begin{figure}
  \includegraphics[width=6cm, angle=-90]{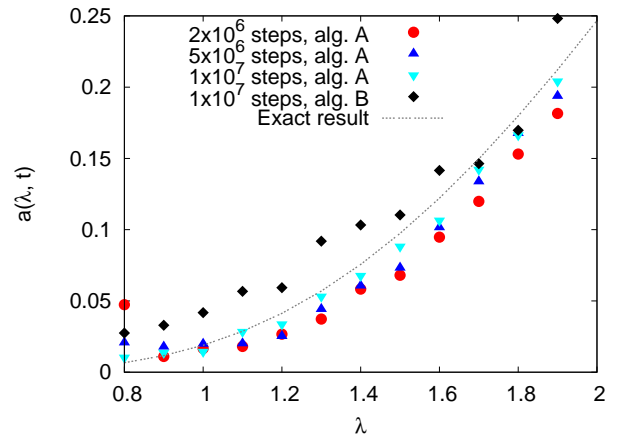}\\
  \caption{Numerical estimates of the lattice spacing as a function of the coupling constant for the two-dimensional large-$N$ $O\lr{N}$ sigma model, compared with the exact result. The estimates were obtained using both algorithms ``A'' and ``B'' with different number of time steps.}
  \label{fig:spacing_vs_lambda}
\end{figure}

 In order to illustrate such a stochastic solution of the equations (\ref{sigma_model_sd}), we consider the case $D = 2$. In two dimensions the model (\ref{sigma_model_pf}) is asymptotically free, and one can introduce the lattice spacing by fixing the value of mass in physical units (we set $m_{phys} = 1$): $m_{phys} \, a\lr{\lambda} = m_{latt}\lr{\lambda} = \sqrt{\kappa^{-1}\lr{\lambda} - 2 D}$. The process of convergence of the lattice spacing to its exact value is illustrated on Fig. \ref{fig:sigma_model_convergence} for both the algorithms ``A'' and ``B''. For algorithm ``A'' we have used $T = 5 \cdot 10^5$. The algorithm ``A'' converges much faster than the algorithm ``B''. The values of lattice spacing obtained using both algorithms are compared with the exact solution on Fig. \ref{fig:spacing_vs_lambda}. In agreement with asymptotic freedom, lattice spacing quickly decreases with $\lambda$. Again, algorithm ``A'' yields more precise results in the same number of time steps.

\section*{Discussion and conclusions}
\label{sec:conclusions}

 In this paper we have presented numerical strategies for the stochastic summation and re-summation of perturbative expansions in large-$N$ quantum field theories. Our basic approach was to interpret the Schwinger-Dyson equations as the equations for the stationary probability distribution of some random process. Since Schwinger-Dyson equations in such theories are nonlinear equations, we had to use so-called nonlinear random processes, rather than ordinary Markov processes whose stationary probability distributions always obey linear equations.

 It is interesting to note that since the configuration spaces of random processes described in this paper are discrete, their numerical implementation require floating-point operations only for the random choice of actions. Thus such algorithms can be potentially much faster than the standard Monte-Carlo simulations based on floating-point arithmetic, and can be advantageous for machines based on GPUs.

 Our final goal is to extend the presented approach to non-Abelian lattice gauge theories. However, in this case direct stochastic interpretation of Schwinger-Dyson equations is only possible at strong coupling, while the continuum limit of such theories corresponds to the weak-coupling limit. In Section \ref{sec:rps_with_mem}, we have discussed a way to access the weak-coupling limit, which, however, was implemented numerically only for $O\lr{N}$ sigma-model at large $N$. The basic idea is to absorb the divergences into a self-consistent redefinition of the expansion parameter and solve the self-consistency conditions using random processes with memory. In some sense, $O\lr{N}$ sigma-model can be thought of as the bosonic random walk in its own condensate, and the approach to the self-consistent value of mass gap (see Fig. \ref{fig:sigma_model_convergence}) - as a renormalization-group flow.

 For non-Abelian gauge theories the redefined expansion parameters can emerge as the lagrange multipliers for the ``zigzag symmetry'' of the QCD string and should also satisfy some self-consistency conditions \cite{Kazakov:94:1}. Zigzag symmetry means that when one adds a line which is passed forward and backward to the boundary of the fluctuating string, the amplitudes should not change. In lattice gauge theory, this condition is equivalent to the unitarity of the link variables $U\lr{x, \mu}$, which is similar to the condition $|n\lr{x}| = 1$ in $O\lr{N}$ sigma-model. These redefined parameters can be also related to the gluon condensate \cite{Kazakov:94:1}. By analogy with the sigma-model, one can think that non-Abelian gauge theories are similar to strings moving in some self-consistent condensates. Such a picture is also close to the idea of holographic AdS/CFT duality for non-Abelian gauge theories, where the dual string lives in some self-consistent gravitational background, and the parameters of this background can be related to gluon condensates in gauge theory \cite{Polyakov:99:1}. In fact, the requirement that the metric of the holographic background approaches that of the AdS space-time ensures the zigzag symmetry of the strings which end on the AdS boundary \cite{Polyakov:99:1}.

 In view of these qualitative considerations, our hope is that the loop equations in non-Abelian gauge theories can be solved stochastically by a random process similar to the one which was devised for the Weingarten model of random surfaces (see Subsection \ref{subsec:weingarten}), but with some self-consistent choice of parameters, which might be implemented as the ``memory'' in the random process.

 Among other possible applications of the presented method one can think of the solution of Schwinger-Dyson equations in continuum gauge theories, combined with the Renormalization Group methods \cite{Pawlowski:07:1}, numerical analysis of quantum gravity models described by various matrix models, and numerical solution of hydrodynamical equations \cite{Migdal:94:2}.

 It should be noted here that several attempts at the stochastic solution of the loop equations in large-$N$ gauge theories have been already described in the literature quite a long time ago \cite{Migdal:86:1}. These algorithms were, in essence, based on the so-called branching random processes, so that the Wilson loop $W\lr{C}$ is proportional to the probability of transition from the initial loop configuration $C$ to the empty configuration with no loops. In particular, in contrast to the algorithm described in Subsection \ref{subsec:weingarten}, where one of the basic steps is to join loops, in the algorithms described in \cite{Migdal:86:1} the basic step was to split a self-intersecting loop into two loops. As a result, these algorithms did not implement the importance sampling and were not able to produce any sensible results for the four-dimensional gauge theory. Generally, branching random processes similar to those considered in \cite{Migdal:86:1} can be obtained from the ``recursive'' nonlinear random process described in this paper by time reversal. However, since such processes do not satisfy any detailed balance condition, they are not invariant under this operation, and lead to very different numerical algorithms.

\begin{acknowledgments}
 I am grateful to Drs. M. I. Polikarpov, Yu. M. Makeenko, A. S. Gorsky, N. V. Prokof'ev and I. Ya. Aref'eva for interesting and stimulating discussions. I'd like also to thank Drs. F. Bruckmann and A. Schaefer for their kind hospitality at the University of Regensburg, where a part of this work was written. This work was partly supported by grants RFBR 09-02-00338-a, RFBR 08-02-00661-a, a grant for the leading scientific schools NSh-6260.2010.2, by the Federal Special-Purpose Programme ``Personnel'' of the Russian Ministry of Science and Education, and by personal grants from the ``Dynasty'' foundation and from the FAIR-Russia Research Center (FRRC).
\end{acknowledgments}

\appendix

\section{Stochastic solution of nonlinear equations with coefficients of arbitrary sign}
\label{sec:arbitrary_coefficients}

 The random process described in Section \ref{sec:recursive_process} was devised under the assumption that the coefficients $p_{c}\lr{x}$, $p_{e}\lr{x|y_1}$ and $p_{j}\lr{x | y_1, y_2}$ are real and positive for any $x$, $y_1$ and $y_2$. In this Appendix we show how the solution of the equation (\ref{random_process_eq}) with arbitrary signs or complex phases on the r.h.s. can be reduced to the solution of another equation of the form (\ref{random_process_eq}) with all positive coefficients, provided the inequalities (\ref{probability_ineq}) are satisfied. We begin by discussing the case of real but non-positive coefficients in details, and finally sketch the extension to complex-valued coefficients.

 To this end, let us represent the sign-alternating coefficients in (\ref{random_process_eq}) as:
\begin{eqnarray}
\label{sign_alt_repr}
p_c\lr{x} = p_c^{\lr{+}}\lr{x} - p_c^{\lr{-}}\lr{x}
\nonumber \\
p_e\lr{x|y} = p_e^{\lr{+}}\lr{x|y} - p_e^{\lr{-}}\lr{x|y}
\nonumber \\
p_j\lr{x|y} = p_j^{\lr{+}}\lr{x|y_1, y_2} - p_j^{\lr{-}}\lr{x|y_1, y_2}  ,
\end{eqnarray}
where $p_c^{\lr{\pm}}\lr{x}$, $p_e^{\lr{\pm}}\lr{x|y}$ and $p_j^{\lr{\pm}}\lr{x|y_1, y_2}$ are all positive and also obey the following inequality:
\begin{eqnarray}
\label{probability_ineq_sign}
\sum \limits_{s = \pm} \sum \limits_{x} p_c^{\lr{s}}\lr{x} + p_e^{\lr{s}}\lr{x|y_1} + p_j^{\lr{s}}\lr{x | y_1, y_2} < 1
\end{eqnarray}
for any $y_1$, $y_2$. Obviously, these inequalities can be satisfied if the inequalities (\ref{probability_ineq}) are satisfied. Let us now introduce two functions $w^{\lr{+}}\lr{x}$ and $w^{\lr{-}}\lr{x}$, which satisfy the following equations:
\begin{widetext}
\begin{eqnarray}
\label{sign_alt_repr_eq}
w^{\lr{+}}\lr{x} = p_c^{\lr{+}}\lr{x} + \sum \limits_{y} p_e^{\lr{+}}\lr{x | y} w^{\lr{+}}\lr{y}
+
\sum \limits_{y} p_e^{\lr{-}}\lr{x | y} w^{\lr{-}}\lr{y}
 + \nonumber \\ +
\sum \limits_{y_1, y_2} p_j^{\lr{+}}\lr{x | y_1, y_2} w^{\lr{+}}\lr{y_1} w^{\lr{+}}\lr{y_2}
+
\sum \limits_{y_1, y_2} p_j^{\lr{+}}\lr{x | y_1, y_2} w^{\lr{-}}\lr{y_1} w^{\lr{-}}\lr{y_2}
+ \nonumber \\ +
\sum \limits_{y_1, y_2} p_j^{\lr{-}}\lr{x | y_1, y_2} w^{\lr{+}}\lr{y_1} w^{\lr{-}}\lr{y_2}
+
\sum \limits_{y_1, y_2} p_j^{\lr{-}}\lr{x | y_1, y_2} w^{\lr{-}}\lr{y_1} w^{\lr{+}}\lr{y_2}
\nonumber \\
w^{\lr{-}}\lr{x} = p_c^{\lr{-}}\lr{x} + \sum \limits_{y} p_e^{\lr{+}}\lr{x | y} w^{\lr{-}}\lr{y}
+
\sum \limits_{y} p_e^{\lr{-}}\lr{x | y} w^{\lr{+}}\lr{y}
 + \nonumber \\ +
\sum \limits_{y_1, y_2} p_j^{\lr{+}}\lr{x | y_1, y_2} w^{\lr{+}}\lr{y_1} w^{\lr{-}}\lr{y_2}
+
\sum \limits_{y_1, y_2} p_j^{\lr{+}}\lr{x | y_1, y_2} w^{\lr{-}}\lr{y_1} w^{\lr{+}}\lr{y_2}
+ \nonumber \\ +
\sum \limits_{y_1, y_2} p_j^{\lr{-}}\lr{x | y_1, y_2} w^{\lr{+}}\lr{y_1} w^{\lr{+}}\lr{y_2}
+
\sum \limits_{y_1, y_2} p_j^{\lr{-}}\lr{x | y_1, y_2} w^{\lr{-}}\lr{y_1} w^{\lr{-}}\lr{y_2}
\end{eqnarray}
\end{widetext}
It is now easy to check that the difference
\begin{eqnarray}
\label{reweighted_w}
w\lr{x} = w^{\lr{+}}\lr{x} - w^{\lr{-}}\lr{x}
\end{eqnarray}
satisfies the equation (\ref{random_process_eq}). On the other hand, the equations (\ref{sign_alt_repr_eq}) again have the form of (\ref{random_process_eq}) with all positive coefficients, but with the configuration space $X'$ being the direct product $X \otimes \mathbb{Z}_2$. In other words, each variable $x$ now in addition carries the ``sign'', $+$ or $-$, which can be written as $\lrc{x, +}$ or $\lrc{x, -}$. Basing on the results presented in Section \ref{sec:recursive_process}, one can devise the random process which solves these equations:
\begin{description}
  \item[Create:] With probability $p_c^{\lr{\pm}}\lr{x}$ create new element $\lrc{x, \pm} \in X'$ and push it to the stack.
  \item[Evolve:] With probability $p_e^{\lr{+}}\lr{x | y}$ pop the element $\lrc{y, \pm}$ from the stack and push the element $\lrc{x, \pm}$ to the stack, with probability $p_e^{\lr{-}}\lr{x | y}$ do the same but flip the sign of $y$.
  \item[Join:] With probability $p_j^{\lr{+}}\lr{x | y_1, y_2}$ consecutively pop two elements $\lrc{y_1, s_1}$, $\lrc{y_2, s_2}$ from the stack and push a single element $\lrc{x, s_1 s_2}$ to the stack. That is, two pluses or two minuses associated with $y$'s give $\lrc{x, +}$, but one plus and one minus give $\lrc{x, -}$. With probability $p_j^{\lr{-}}\lr{x | y_1, y_2}$ do the same, but flip the resulting sign, that is, push the element $\lrc{x, - s_1 s_2}$ to the stack.
  \item[Restart:] Otherwise, empty the stack and push a single element $\lrc{x, \pm} \in X'$ into it, with probability proportional to $p_c^{\lr{\pm}}\lr{x}$.
\end{description}
As in Section \ref{sec:recursive_process}, $w^{\lr{+}}\lr{x}$ and $w^{\lr{-}}\lr{x}$ are proportional to probabilities to find the elements $\lrc{x, +}$ or $\lrc{x, -}$ on the top of the stack, provided there is more than one element in it.

 The extension of this construction to complex-valued coefficients is quite straightforward. We represent the coefficients in (\ref{random_process_eq}) as $p_c\lr{x} = \int \limits_{0}^{2 \pi} d\theta \, p_c\lr{x, \theta} \expa{i \theta}$, where $p_c\lr{x, \theta}$ is real and positive, and similarly for the other coefficients. The configuration space $X'$ becomes the direct product $X \otimes S^1$, where $S^1$ is the unit circle in the complex plane. The stack now contains the pairs $\lrc{x, \theta}$, with $\theta \in \lrs{0, 2 \pi}$ being the complex phase. The function $w\lr{x}$ is estimated as $w\lr{x} =  \int \limits_{0}^{2 \pi} d\theta \, w\lr{x, \theta} \expa{i \theta}$, where $w\lr{x, \theta}$ is the probability to find the element $\lrc{x, \theta}$ at the top of the stack. In the random process, new elements are created with probability distribution proportional to $p_c\lr{x, \theta}$, and in the ``Evolve'' and the ``Join'' actions the phases of the elements and the coefficients in (\ref{random_process_eq}) are added modulo $2 \pi$, similarly to signs.

 Note that this solution does not in general have the property of importance sampling. Indeed, one can have some $x$ for which $w\lr{x}$ is numerically very small, but the random process can spend an almost equal large amount of time in the states $\lrc{x, +}$ and $\lrc{x, -}$, so that numerically large $w^{\lr{+}}\lr{x}$ and $w^{\lr{-}}\lr{x}$ nearly cancel. Whether this occurs or not depends on the particular system of equations and on the particular unknown variables, but potentially this feature can make numerical simulations less efficient.

%\bibliography{MyBibliography}
%\bibliographystyle{apsrev}

\end{document}